\documentclass{article}
\PassOptionsToPackage{numbers, sort&compress}{natbib}
\usepackage[preprint]{neurips_2026}

\usepackage[utf8]{inputenc}
\usepackage[T1]{fontenc}
\usepackage{amsmath}
\usepackage{hyperref}
\usepackage{cleveref}
\usepackage{url}
\usepackage{booktabs}
\usepackage{array}
\usepackage{multirow}
\usepackage{graphicx}
\usepackage{caption}
\usepackage{CJKutf8}
\usepackage{amsfonts}
\usepackage{enumitem}
\usepackage{nicefrac}
\usepackage{microtype}
\usepackage{xcolor}
\usepackage{colortbl}
\usepackage{listings}
\usepackage[most]{tcolorbox}
\usepackage{tikz}
\usepackage{float}

\definecolor{codeblue}{RGB}{56,118,189}
\definecolor{codegray}{RGB}{120,120,120}
\definecolor{codegreen}{RGB}{44,140,90}
\definecolor{codeorange}{RGB}{196,98,16}
\definecolor{codepurple}{RGB}{130,80,160}
\definecolor{codebg}{RGB}{252,252,252}
\definecolor{codeframe}{RGB}{190,195,205}
\definecolor{codetitle}{RGB}{62,68,82}
\lstdefinestyle{netcode}{
  backgroundcolor=\color{codebg},
  basicstyle=\ttfamily\scriptsize,
  keywordstyle=\color{codeblue}\bfseries,
  stringstyle=\color{codegreen},
  commentstyle=\color{codegray}\itshape,
  numberstyle=\tiny\color{codegray},
  numbers=left,
  numbersep=5pt,
  xleftmargin=12pt,
  framexleftmargin=12pt,
  breaklines=true,
  showstringspaces=false,
  tabsize=2,
  columns=fullflexible,
  keepspaces=true,
  literate={:}{\textcolor{codeorange}{:}}1
           {,}{\textcolor{codegray}{,}}1
           {[}{\textcolor{codepurple}{[}}1
           {]}{\textcolor{codepurple}{]}}1
           {\{}{\textcolor{codepurple}{\{}}1
           {\}}{\textcolor{codepurple}{\}}}1,
}
\newtcblisting{codebox}[1][]{
  enhanced,
  title={#1},
  fonttitle=\scriptsize\bfseries\sffamily,
  colback=codebg,
  colframe=codeframe,
  coltitle=white,
  colbacktitle=codetitle,
  boxrule=0.3pt,
  arc=1.5pt,
  left=1pt, right=3pt, top=1pt, bottom=1pt,
  toptitle=1.5pt, bottomtitle=1.5pt,
  listing only,
  listing options={style=netcode},
}
\usetikzlibrary{arrows.meta,calc,positioning,fit,backgrounds,matrix,shapes.geometric}
\crefname{section}{}{\S\S}
\crefdefaultlabelformat{\S#2#1#3}

\newcommand{\methodnaive}{{In-Context Learning}}
\newcommand{\methodmulti}{{Iterative Reasoning}}
\newcommand{\methodagent}{{Agent Skills}}
\newcommand{\benchmarkname}{\textsc{RoutingBench}}

\definecolor{accGood}{RGB}{160,210,160}
\definecolor{accGoodMid}{RGB}{175,217,175}
\definecolor{accMid}{RGB}{195,225,195}
\definecolor{accLowMid}{RGB}{215,235,215}
\definecolor{accLow}{RGB}{235,246,235}
\definecolor{accVLow}{RGB}{248,252,248}
\definecolor{tokVLow}{RGB}{254,242,242}
\definecolor{tokLow}{RGB}{251,233,233}
\definecolor{tokMid}{RGB}{240,195,195}
\definecolor{tokHi}{RGB}{228,170,170}
\definecolor{rndVLow}{RGB}{255,247,228}
\definecolor{rndLow}{RGB}{254,238,210}
\definecolor{rndMid}{RGB}{242,195,130}
\definecolor{rndHi}{RGB}{235,180,110}
\definecolor{rtVLow}{RGB}{248,242,251}
\definecolor{rtLow}{RGB}{225,210,240}
\definecolor{rtMid}{RGB}{205,180,228}
\definecolor{rtHi}{RGB}{190,165,220}
\definecolor{tlVLow}{RGB}{255,250,225}
\definecolor{tlLow}{RGB}{252,240,185}
\definecolor{tlMid}{RGB}{248,225,140}
\definecolor{tlHi}{RGB}{240,205,95}
\definecolor{rtoLow}{RGB}{228,242,240}
\definecolor{rtoMid}{RGB}{180,220,215}
\definecolor{rtoHi}{RGB}{130,195,190}
\newcommand{\accH}[1]{\cellcolor{accGood}#1}
\newcommand{\accHM}[1]{\cellcolor{accGoodMid}#1}
\newcommand{\accM}[1]{\cellcolor{accMid}#1}
\newcommand{\accLM}[1]{\cellcolor{accLowMid}#1}
\newcommand{\accL}[1]{\cellcolor{accLow}#1}
\newcommand{\accVL}[1]{\cellcolor{accVLow}#1}
\newcommand{\tokVL}[1]{\cellcolor{tokVLow}#1}
\newcommand{\tokL}[1]{\cellcolor{tokLow}#1}
\newcommand{\tokM}[1]{\cellcolor{tokMid}#1}
\newcommand{\tokH}[1]{\cellcolor{tokHi}#1}
\newcommand{\rndVL}[1]{\cellcolor{rndVLow}#1}
\newcommand{\rndL}[1]{\cellcolor{rndLow}#1}
\newcommand{\rndM}[1]{\cellcolor{rndMid}#1}
\newcommand{\rndH}[1]{\cellcolor{rndHi}#1}
\newcommand{\rtVL}[1]{\cellcolor{rtVLow}#1}
\newcommand{\rtL}[1]{\cellcolor{rtLow}#1}
\newcommand{\rtM}[1]{\cellcolor{rtMid}#1}
\newcommand{\rtH}[1]{\cellcolor{rtHi}#1}
\newcommand{\tlVL}[1]{\cellcolor{tlVLow}#1}
\newcommand{\tlL}[1]{\cellcolor{tlLow}#1}
\newcommand{\tlM}[1]{\cellcolor{tlMid}#1}
\newcommand{\tlH}[1]{\cellcolor{tlHi}#1}
\newcommand{\rtoL}[1]{\cellcolor{rtoLow}#1}
\newcommand{\rtoM}[1]{\cellcolor{rtoMid}#1}
\newcommand{\rtoH}[1]{\cellcolor{rtoHi}#1}
\definecolor{accTextA}{RGB}{0,0,0}
\definecolor{accTextB}{RGB}{0,0,0}
\definecolor{accTextC}{RGB}{0,0,0}
\definecolor{accTextD}{RGB}{0,0,0}

\title{\benchmarkname: Can Agentic Routing Analysis\\ Scale to Production Datacenter Networks?}
\author{
  Wenlong Ding$^{1,2}$, Zhixiong Niu$^{2}$, Jianan Yang$^{3}$, Fajun Zhang$^{3}$, Bo Zhang$^{3}$ \\
  \bf Ling Liang$^{3}$, Yongqiang Xiong$^{2}$, Tianyin Xu$^{4}$, Hong Xu$^{1}$ \vspace{3pt} \\
  $^{1}$CUHK \quad $^{2}$Microsoft Research Asia \quad $^{3}$Microsoft \quad $^{4}$UIUC
}

\begin{document}

\begin{CJK*}{UTF8}{gbsn}

\maketitle

\begin{abstract}
Recent advances in AI models and agentic technologies 
    make AI for network operations (NetOps) within reach.
However, scalability remains a key bottleneck of agentic NetOps 
    when analyzing hyperscale networks, 
    which comprise hundreds of datacenters,
    each housing thousands of network devices.
The scalability challenge is rooted in the requirement of many NetOps tasks 
    that must conduct {\it global} reasoning on how a {\it local} change of 
    device behavior affects {\it all} relevant routing paths,
    known as {\it routing-path analysis}.
This paper studies this scalability problem and evaluates how different agentic approaches, 
    namely in-context learning, iterative reasoning, and agent skills, 
    can scale routing-path analysis to large, complex networks.
We present \benchmarkname{} for evaluating agentic routing-path analysis, 
    with varying network size and complexity, for various types of device changes.
Our results show that agentic analysis is promising---agent 
    skills curated with a principle termed ``\emph{explore more; digest less}'' 
    enables path analysis on hyperscale networks 
    of 50K routers with an accuracy of 99.5\%,
    significantly outreaching the scalability of traditional symbolic analysis.
Meanwhile, \benchmarkname{} reveals the boundary of AI agent capability 
    on complex inter-datacenter networks and compound changes, 
    posing open challenges for AI and agentic research.
\end{abstract}

\section{Introduction}
\label{sec:intro}

With recent advances in model capabilities and agentic technologies,
    AI for Network Operations (NetOps) has been increasingly explored to automate various tasks 
    such as command/script generation, capacity planning, monitoring data analytics, etc~\cite{confucius,NetLLM,xumi,netconfeval}.
However, scalability remains a key bottleneck to applying AI for NetOps 
    in hyperscale networks~\cite{azure-dc,azure,s2,batfish2}.
To give a concrete data point, the hyperscale network we manage 
    connects 300+ geo-distributed data centers, each housing more than 10K network devices;
    each device is instructed by complex configuration files with more than 5K lines,
    specifying runtime behavior of the devices based on various network protocols.
How to understand the behavior of such massive-scale networks, especially upon (planned or unexpected) changes, 
    has been one of the grand challenges of NetOps research~\cite{dna,rela,relationalnetkat}.

The scalability challenge is rooted in the requirement of many NetOps tasks 
    (e.g., for reachability and security analysis) 
    that must conduct {\it global} reasoning on how a {\it local} change of device behavior (e.g., protocol configurations)
    affects {\it all} relevant routing paths.
We refer to the core primitive of these tasks as {\it routing-path analysis} (or {\it path analysis} in short),
    which must reason about 
    end-to-end, network-wide paths based on the network topology and device configurations.
For example, to analyze the impact of a device-configuration change (in terms of safety and security),
    path analysis is used to compare the paths before and after the change~\cite{HSA,batfish,batfish2}.
Path analysis is inherently expensive to scale, 
    because it requires {\it cross-device} and {\it cross-protocol} reasoning, e.g., 
    changing BGP (Border Gateway Protocol)'s Local Preference (LP)~\cite{junos-bgp-local-preference} 
    for a single IP (Internet Protocol) prefix on one router could redirect traffic to a different next hop, 
    which in turn changes all downstream protocols and policies across routers, 
    creating a combinatorial reasoning space over possible paths.

Traditionally, path analysis is conducted by symbolic analysis~\cite{batfish,batfish2,Minesweeper,ARC,NetDice,QARC,NDD,dna},
    which needs to construct whole-network routing tables (typically in a mathematical representation).
The scalability of these tools is thus limited by the time and memory costs which 
    scale poorly with the size and complexity of the network.
Batfish~\cite{batfish2}, a state-of-the-art analysis tool, takes hours to analyze 
    paths for a given pair of IP prefixes of a network with 5K devices and runs out of memory beyond 
    8K devices on a powerful server with one terabyte of physical memory (see \Cref{sec:rationale}).
Due to such scalability bottleneck, it is hard to completely (or even comprehensively) analyze the safety 
    and security properties of device changes, even though they have been 
    the dominating causes of disruptions and outages of production networks~\cite{fb-outage-2021,azure-outage-2023,google-outage-2020}.

This paper explores whether {\it agentic} approaches can address the long-lasting scalability bottlenecks of routing-path analysis.
Large language models (LLMs) are pretrained with domain knowledge 
    of network protocols (e.g., TCP/IP and BGP) and fluent in configuration
    languages~\cite{protocol-understand,topo-understand,topo-understand2,intent-extract,cosynth,cegs,netcomplete};
agent technologies further enable multi-step reasoning over long context~\cite{chainofthought,rag,skills}. 
In the BGP LP example above, an agent could iteratively extract and reason over relevant configuration evidence 
    (e.g., IP prefixes, static routes, BGP sessions) across devices to identify affected paths.
The key aspect we study is not LLM literacy on network configurations, 
    but how an agent can efficiently discover decisive configuration evidence 
    under context pressure and long reasoning chains at scale.

We thus present \benchmarkname{}, the first benchmark which challenges AI agents to conduct path analysis 
    for different scenarios of device changes
    in networks at varying scales.
\benchmarkname{} automatically generates realistic device configurations 
    for widely used datacenter topologies~\cite{batfish2,dna,fattree1,fattree2,azure-dc,jupiter,fb-fabric}, 
    modeled after production practices---a network-protocol layer (e.g., IP, BGP) 
    ensures full endpoint connectivity, while routing-control configurations (e.g., BGP LP, static routes) 
    govern specific flows.
\benchmarkname{} then measures accuracy, scalability, and cost under a unified evaluation framework, 
    ensuring fair comparison across agentic approaches and LLMs.

We use \benchmarkname{} to evaluate three agentic approaches on routing-path analysis, 
    namely {\it \methodnaive{}}, {\it \methodmulti{}}, and {\it \methodagent{}}, 
    which represent different tradeoffs between context and planning pressures.
On one end, \methodnaive{} includes the network topology and all raw configurations in one prompt, 
    so both evidence retrieval and path reasoning happen entirely in the context window;
on the other end, \methodagent{} starts with no topology or configurations in context 
    and retrieves what it needs through iterative plan-extract-reason loops 
    over routing-specific skills~\cite{skills}. 
In between, \methodmulti{} provides the full topology without configurations, 
    letting the agent choose which devices to inspect and read relevant configurations over multiple rounds.

Our results show that \methodagent{} with general-purpose LLMs (e.g., GPT-5.4) 
    outperforms both other agentic approaches and thinking models.
Specifically, GPT-5.4 with \methodagent{} scales to 50K routers with 99.5\% accuracy and {<}120K 
    tokens within an hour;
in comparison, \methodnaive{} and \methodmulti{} overflow the context window 
    at 500 and 3.1K routers, respectively.
Meanwhile, we show that in complex scenarios (such as inter-datacenter topologies and compound changes), performance of GPT-5.4 with \methodagent{}
    degrades from 90\%+ at dozens of routers to 70\%+ at thousands of routers, marking the current capability boundary of agentic path analysis.
\benchmarkname{} serves as a foundation to prepare agentic NetOps technologies for production uses.

\section{Background}
\label{sec:rationale}

The goal of network analysis is to understand end-to-end routing behavior
  based on configurations of network devices.
The essential challenge comes from the scale of real-world networks.
For example,
our hyperscale network infrastructure contains 300+ interconnected datacenters, each
  housing 10K+ routers (the largest one has 20K routers).
Configuration changes of network devices occur on an hourly or a daily basis depending 
  on the types (Table~\ref{tab:Configupdate}).
The most frequent changes include adding or removing a network device and updating 
  the IP prefixes of a device (e.g., to deploy new policies).

\begin{figure}[t]
\centering
\includegraphics[width=\linewidth]{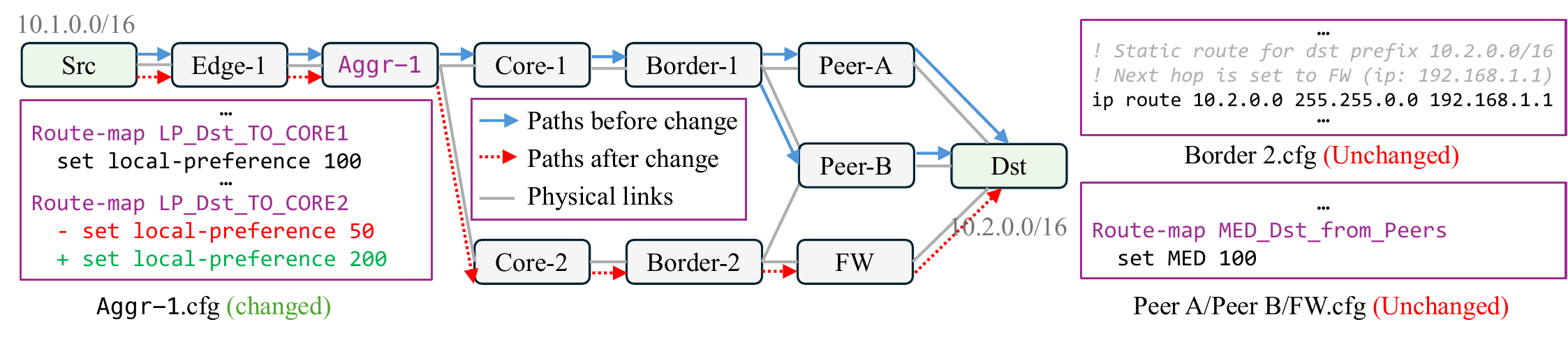}
\vspace{-18pt}
\caption{An example of routing-path analysis.
In this example, a local BGP LP change on router \texttt{Aggr-1} would direct the queried endpoint-prefix pair 
  onto a globally different path. Along the new path, determining the path outcome requires reasoning across 
  network devices and protocols.}
\label{fig:path-analysis-example}
\vspace{-10pt}
\end{figure}

Many network analysis tasks can be reduced to path analysis which determines
  how the network forwards a target {\it flow} under a given network state 
  (defined by the configurations of all the devices in the network).
A network flow is specified by (1) source and destination IP prefixes, and
  (2) ingress (where traffic enters the network), and the egress (where traffic exits) routers.
For a given flow,
  path analysis asks which sequence of routers it would traverse across the network. 
Figure~\ref{fig:path-analysis-example} shows an example, where the flow is 
  \texttt{<10.1.0.0/16,} \texttt{10.2.0.0/16,} \texttt{Src,} \texttt{Dst>}.
When a router configuration changes (e.g., the BGP LP change in \texttt{aggr-1.cfg}), 
  routing paths of the flow 
  change accordingly, which could affect properties like reachability, security, load balance, etc.

\begin{figure}[t]
\centering
\begin{minipage}[t]{0.48\linewidth}
\centering
\includegraphics[width=0.9\linewidth]{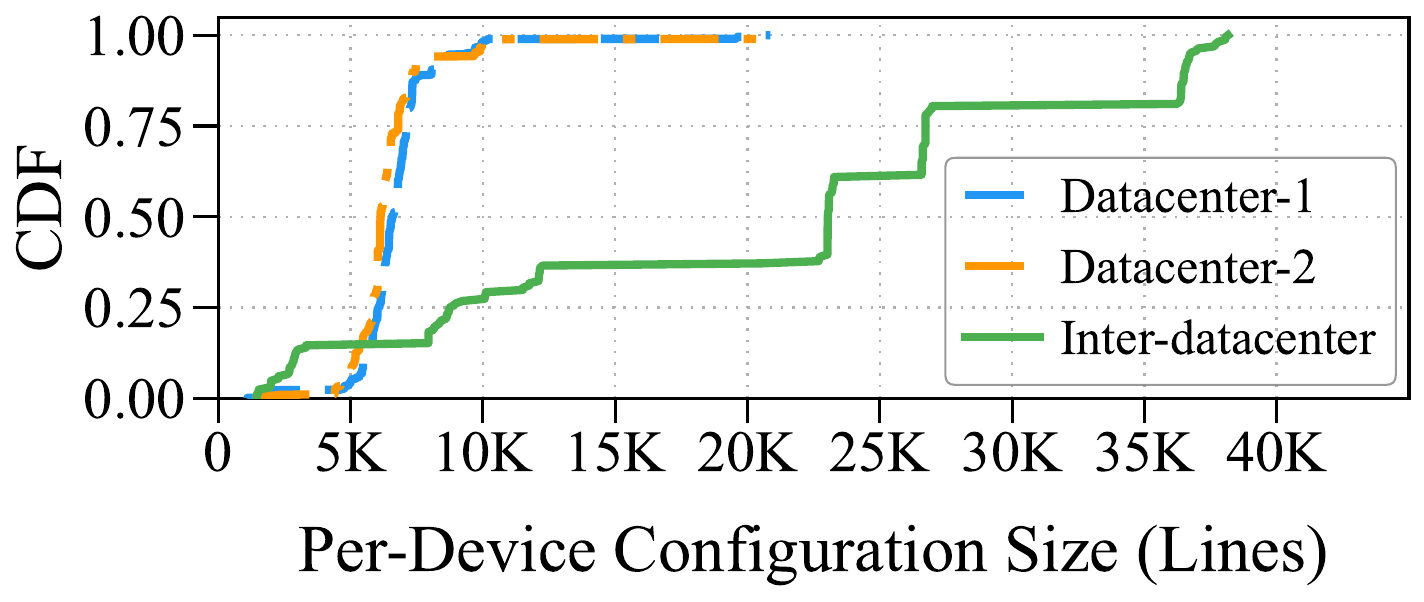}
\vspace{-5pt}
\caption{CDF of per-device configuration size (in lines) for the routers within 
  two production datacenters and inter-datacenter routers.}
\label{fig:config-cdf}
\vspace{-20pt}
\end{minipage}\hfill
\begin{minipage}[t]{0.48\linewidth}
\centering
\includegraphics[width=0.9\linewidth]{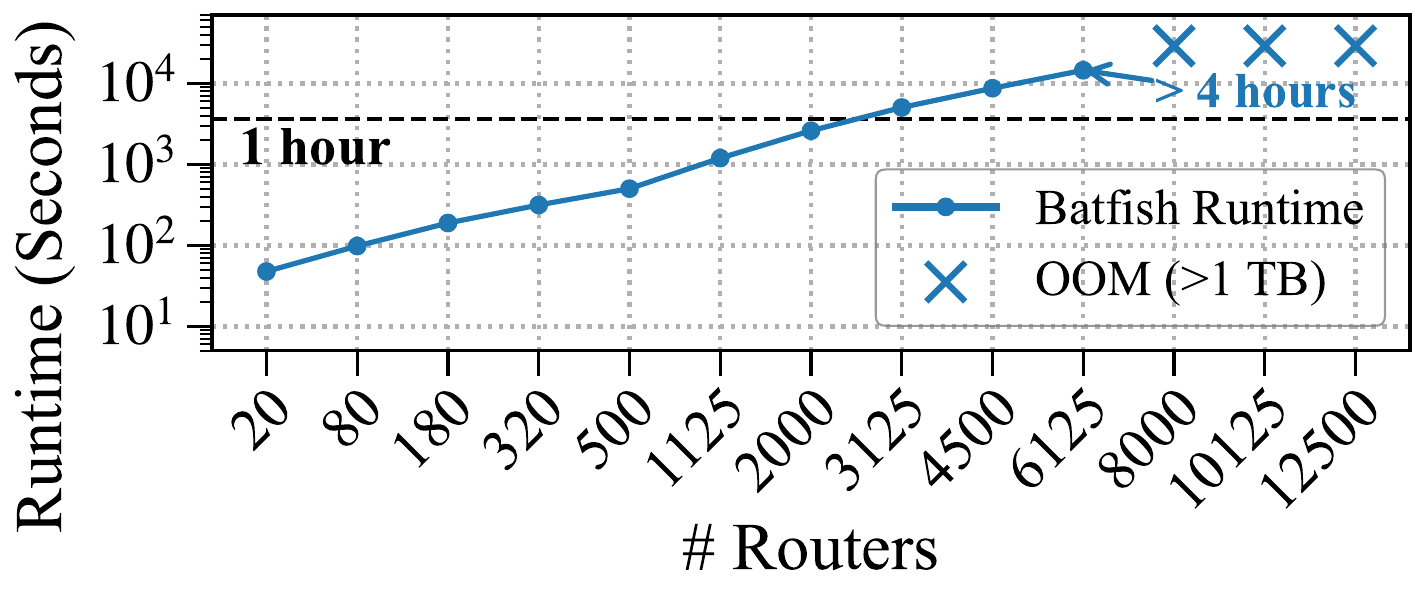}
\captionsetup{type=figure}
\vspace{-5pt}
\caption{Batfish runtime across topology sizes. 
  Neither runtime nor memory can scale to a single production-size datacenter 
  with 10K+ routers.}
\label{fig:batfish-runtime}
\vspace{-20pt}
\end{minipage}
\end{figure}

However, path analysis is intrinsically difficult for two reasons. 
First, it is not a \textit{local} task: the effect of a configuration change rarely remains confined to 
  the device where it is made. 
As shown in Figure~\ref{fig:path-analysis-example}, a local BGP LP change on an aggregation router (\texttt{Aggr-1}) 
  can redirect traffic from one path to a completely different path.
Second, path analysis is inherently \textit{cross-protocol} and \textit{cross-device}. 
In Figure~\ref{fig:path-analysis-example}, determining Border~2's next hop requires the analysis 
  to identify relevant routing configurations across Border~2, Peer~B, and FW, 
  including static routes~\cite{junos-static-routes} that explicitly set the next hop for a prefix 
  and BGP MED~\cite{junos-bgp-med} that ranks alternative BGP peers with lower values preferred. 
An effective path analysis must analyze configurations across multiple devices 
  and understand that the static route determines 
  the path because it has higher priority than BGP MED in terms of protocols. 
Note that device configurations in production networks are complex and large in size.
Figure~\ref{fig:config-cdf} shows configuration-file-size distributions of 
  intra- and inter-datacenter routers.
Within a datacenter, 95+\% of devices exceed 5K lines to
  specify interfaces, prefix lists, route maps, BGP policies, etc.
Inter-datacenter routers are substantially larger: more than 70\% exceed 10K lines, 
  with the largest reaching 38K lines; such routers like gateways 
  encode additional state, broader prefix-filtering policies, and more route-map entries to accommodate the larger number of connected interfaces and traffic classes.

{\bf Scalability bottleneck.} Today, path analysis is done through static analysis which
  checks whether routing configurations induce intended forwarding outcomes~\cite{batfish,batfish2,Minesweeper,ARC,NetDice,QARC,ProbNetKAT}.
For a given flow, these static analysis tools return the routing paths.
To do so, these tools construct routing graphs and routing tables for each device by parsing and modeling 
  network-wide device configurations and topology. 
For example, Batfish~\cite{batfish,batfish2} uses simulation to derive this routing information 
  after modeling the network. The tools then answer end-to-end queries by traversing those graphs 
  and tables.
Such whole-network model enables accurate path analysis, but creates scalability bottlenecks---the analysis complexity 
  increases with the number of devices, prefixes, and policy interactions.

We quantify the performance bottleneck of existing analysis tools using a set of controlled experiments.
Figure~\ref{fig:batfish-runtime} shows the time it takes for Batfish~\cite{batfish,batfish2}
  to analyze a given network flow.
The analysis takes more than one hour for a network with 3K routers 
  and more than four hours for 6K routers.
It runs out of memory for a network with 8K routers on a powerful server with one terabyte physical memory.
So, the scalability of Batfish is insufficient even for individual datacenters,
  let alone inter-datacenter networks. 
In current practices, NetOps engineers work around this bottleneck by selecting 
  a subset of devices based on heuristics or experience (e.g., certain clusters or gateway routers). 

\section{\benchmarkname{}}
\label{sec:benchmark}

We build \benchmarkname{} to systematically evaluate AI agents on routing-path analysis at scale. 
\benchmarkname{} covers both simple cases (e.g., individual configuration changes 
    within a single datacenter) and more complex ones
    (e.g., compound configuration changes and inter-datacenter setups).
Specifically, \benchmarkname{} implements a {\it fully automatic} benchmark generation 
    framework rather than a fixed collection of hand-picked cases: 
    it generates network topologies (Clos networks~\cite{fattree1,fattree2}) 
    and matched device configurations, synthesizes path-analysis queries with one or more configuration changes, 
    and evaluates NetOps agents through unified interfaces and metrics.
\begin{figure}[H]
\centering
\begin{minipage}[t]{0.615\textwidth}
  \centering
  \includegraphics[width=\linewidth]{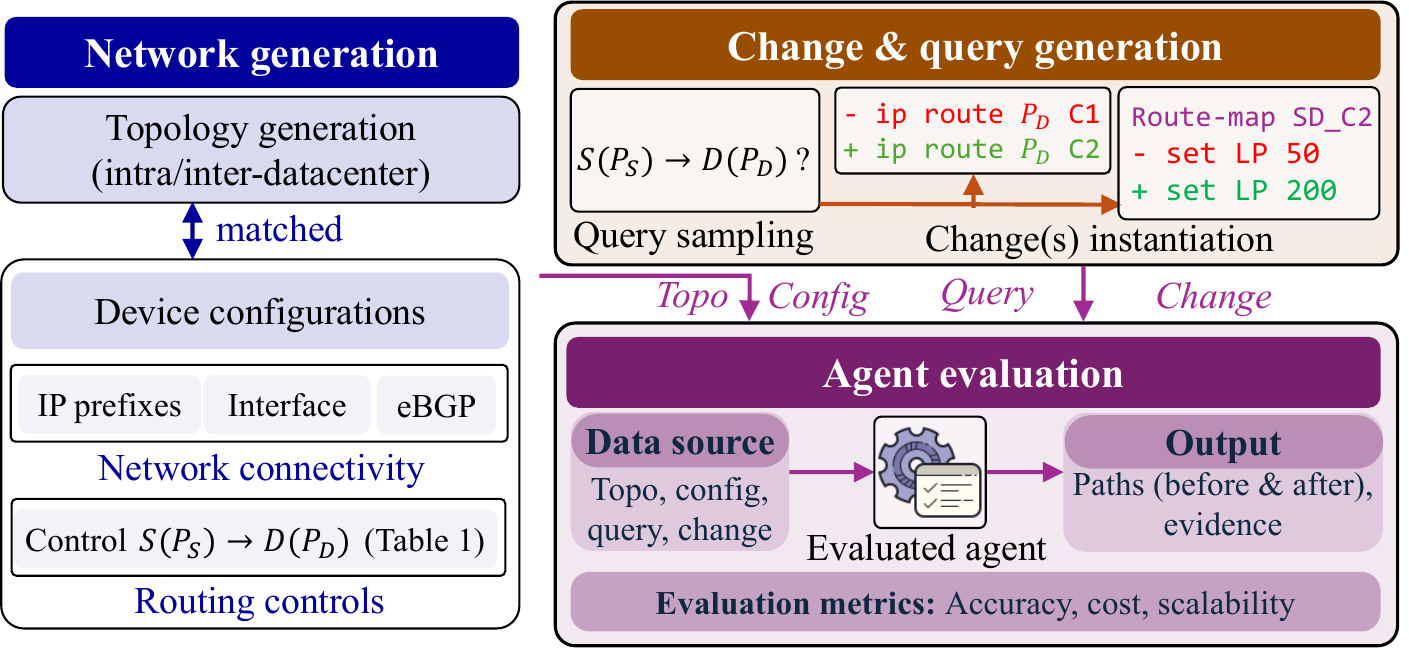}
  \vspace{-12pt}
  \caption{{\bf \benchmarkname{} construction overview.} 
    It defines a shared path-analysis task with unified metrics and automatically generates hyperscale 
    network instances with configuration changes at varying topology scales.}
  \label{fig:benchmark}
\end{minipage}\hfill
\begin{minipage}[t]{0.345\textwidth}
  \centering
  \includegraphics[width=\linewidth]{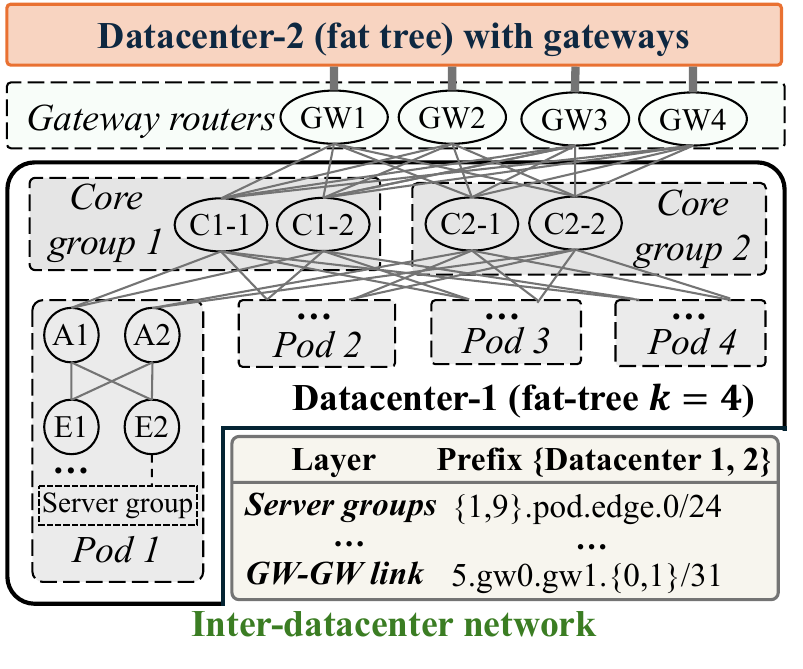}
  \vspace{-12pt}
  \caption{{An example of a generated datacenter network (modeled as a \textit{fat tree}) 
    and conflict-free IP prefix assignment templates.}}
  \label{fig:fattree}
\end{minipage}
  \vspace{-7.5pt}
\end{figure}

\subsection{Task Formulation}
\label{subsec:task}

A \benchmarkname{} task is a path-analysis query: given
    the network topology, configuration files for all devices,
    a target network flow (a source-destination endpoint pair $(S, D)$
    and their corresponding prefix pair $(P_S, P_D)$),
    and one or more configuration changes,
    the agent should produce \emph{all} routing paths before and after the change
    with decisive configuration evidence (Figure~\ref{fig:benchmark}).
Each path is an ordered router sequence like $S \rightarrow A \rightarrow \cdots \rightarrow D$.
This task is the basic primitive of many higher-level NetOps tasks.
\benchmarkname{} randomly selects query pairs $(S, D)$ with $(P_S, P_D)$ from the generated endpoints, 
and also supports user-specified pairs for targeted evaluation.
We focus on configuration changes, which are the most frequent events
    that trigger NetOps analysis (Table~\ref{tab:Configupdate}).
{\bf Individual changes.}
\benchmarkname{} automatically generates different types of configuration changes
    as shown in Table~\ref{tab:advanced-configs}.
These configuration changes are based on our operation experiences and are consistent   
    with prior research on network management~\cite{azure-dc,dna,ARC,batfish2}.
\benchmarkname{} first selects a query pair (endpoints $S$ and $D$ with their prefixes $P_S$ and $P_D$) 
    and a configuration type in Table~\ref{tab:advanced-configs}.
It then localizes the relevant lines that control the specific configuration 
    of each prefix and modifies the target attribute to produce the change
    (e.g., rewriting the next-hop field in a static-route template, or raising the BGP local preference toward a different neighbor 
     as shown in Figure~\ref{fig:benchmark}).
For Interface Shutdown changes,
    it selects a router interface along a path from $S$ to $D$ 
    and appends a shutdown command to that interface configuration.
{\bf Compound changes}.
\benchmarkname{} also generates compound changes of the same flow across protocols, which  
    are common in production NetOps.
It does so by assembling multiple individual changes to alter routing behavior of a flow. 
For example, \benchmarkname{} can change the flow through both the static route next hop and BGP LP (Figure~\ref{fig:benchmark}). 
Specifically, we construct four types of compound changes: 
    LP\,+\,BGP~NS, LP\,+\,Shutdown, LP\,+\,Static~Route\,+\,Shutdown, and LP\,+\,Static~Route\,+\,BGP~NS\,+\,Shutdown.
They cover 2--4 changes and introduce combinatorial complexity 
    to evaluate the ability of NetOps agents to understand interactions between protocols.

\subsection{Network Generation}
\label{subsec:dataset}

\begin{table}[t]
\centering
\caption{
Configuration types used in benchmark tasks. (\Cref{tab:advanced-configs-full} is the detailed version.)}
\label{tab:advanced-configs}
\resizebox{\textwidth}{!}{
\renewcommand{\arraystretch}{1.1}
\begin{tabular}{@{}ll@{}}
\toprule
Configuration type & Effect \\
\midrule
BGP Local Preference (LP)~\cite{cisco-local-preference,junos-bgp-local-preference} & Selects the next-hop neighbor with the highest LP for a flow. \\
Multi-Exit Discriminator (MED)~\cite{cisco-med,junos-bgp-med} & Selects the ingress neighbor with the lowest MED for a flow. \\
Static route~\cite{cisco-static,junos-static-routes} & Pins a flow to a fixed next hop, overriding BGP-learned paths. \\
BGP network advertisement~\cite{cisco-bgp-network,junos-bgp-network} & Adds or removes a prefix from BGP, toggling flow reachability. \\
Prefix aggregation~\cite{cisco-aggregate-route,junos-aggregate-route} & Merges prefixes into a group; all matching flows take the same paths. \\
Interface shutdown~\cite{cisco-interface-shutdown,junos-interface-shutdown} & Disables a link, forcing traffic onto an alternate path. \\
\bottomrule
\end{tabular}}
\end{table}

\textbf{Topology.}
\benchmarkname{} generates intra- and inter-datacenter network topologies.
A datacenter network uses a fat-tree based Clos topology~\cite{fattree2,fattree1}.
The scale of the network is parameterized by an integer~$k$, representing 
    the number of pods where each pod is a group of edge and aggregation routers 
    sharing a set of server endpoints.
As shown in Figure~\ref{fig:fattree}, a datacenter network has three layers: edge, aggregation, and core networks. 
Within each pod, the $k/2$ edge routers and $k/2$ aggregation routers are fully connected.
The core layer contains $k/2$ groups with $k/2$ routers each; every router in a group is connected 
    to the aggregation router with the same index across all pods (e.g., the $i$-th aggregation router 
    in every pod connects to all core routers in group~$i$).
Each edge router connects to $k/2$ servers, giving the datacenter $k^3/4$ servers 
    (i.e., IP prefixes) and $5k^2/4$ routers in total.

For inter-datacenter networks, \benchmarkname{} instantiates two datacenters
    and adds $k$ gateway routers per datacenter above the core layer (Figure~\ref{fig:fattree}).
The gateways are fully connected to their own datacenter's core layer and to all gateways in the other datacenter.
Benchmark queries then place the source and destination edge routers in different datacenter networks.
We currently do not include inter-datacenter routers other than gateways (which are mostly controlled 
    by a software-defined network controller~\cite{onewan,swan} instead of device configurations).
\Cref{app:fattree-gen} contains more implementation details about topology generation.

\textbf{Device configuration.}
\benchmarkname{} then generates device configurations for each router
    in the network.
The configuration has two parts: 
    (1)~\emph{connectivity} of all endpoints (servers) in the networks, 
    and (2)~\emph{routing} between endpoints that match the complexity of real-world networks.

For connectivity, \benchmarkname{} first assigns IP prefixes to endpoints 
    and inter-router links (two interfaces on both sides of a link must share the same subnet) without conflict.
As shown in Figure~\ref{fig:fattree}, within a datacenter, 
    \benchmarkname{} encodes the pod, edge/aggregation/core router IDs into 
    relevant prefixes to enable unique IPs for different devices.
For example, a group of endpoints attached to a certain edge router in a pod is assigned \texttt{1.pod.edge.0/24}, providing 256~IPs.
For inter-router links that require only two IPs (downstream and upstream interfaces), 
    \benchmarkname{} uses \texttt{/31} prefixes.
The most significant byte of each prefix indicates the network layer and enables inter-datacenter 
    addressing (e.g., \texttt{5} denotes gateway-to-gateway links, and \texttt{9} denotes server groups of DC1 in Figure~\ref{fig:fattree}).
Since each field occupies one byte, this scheme supports up to $k=256$ (pods), 
    already enabling far larger networks than production scale ({>}\,80K routers).
\benchmarkname{} widens each field to two bytes using IPv6 prefixes with the same positional encoding for larger topologies.
Once all prefixes are assigned, \benchmarkname{} configures router interfaces by filling 
    the corresponding prefixes into the interface templates.
\benchmarkname{} then generates eBGP~\cite{bgp-cisco} configuration, 
    filling in each device's neighbor interface IPs, 
    and advertised endpoint prefixes so that neighboring devices 
    establish BGP sessions and all flows become reachable. (See~\Cref{app:config-gen} for more details.)

\benchmarkname{}'s configuration generation is topology-agnostic: we use fat trees as dominant datacenter topology, 
but other ones (e.g., leaf-spine~\cite{azure-dc}, dragonfly~\cite{dragonfly}) can be supported similarly.

For routing, \benchmarkname{} generates specific configurations in Table~\ref{tab:advanced-configs} 
    for randomly selected prefix pairs (Figure~\ref{fig:benchmark}). 
\benchmarkname{} uses configuration templates (\Cref{app:ad-config-gen}) for each configuration type 
    and the specific prefixes and router names within the generated network.

\subsection{Metrics}
\label{subsec:metrics}

{\bf Accuracy.}
    We measure correctness by requiring the agent's analysis, in terms of router sequences, 
        to precisely match the ground truth,
    i.e., an analysis is correct if and only if the paths before and after the change are both correct.
The ground truth is obtained by using Batfish~\cite{batfish,batfish2} traceroute simulation 
    on topologies where it completes within memory limits (roughly $\le$8K routers; see \S\ref{fig:batfish-runtime}).
For larger topologies that exceed Batfish's capacity, we manually construct routing-related configurations 
    whose routing effects can be derived analytically, yielding deterministic ground-truth paths.

 {\bf Cost.}
    We report \textit{token consumption} and agent \textit{run time}.
    Token consumption is the sum of input and output tokens across all LLM rounds (\methodmulti{} and \methodagent{} 
        invoke LLMs multiple times).
    Run time includes LLM reasoning and tool invocation.
{\bf Scalability.}
    We evaluate the maximum network scale at which an agent remains effective under two constraints: 
    (1) the total input and output tokens do not exceed the LLM's context window, 
    and (2) a path analysis completes within a fixed time threshold (1~hour in our evaluation) to meet operational expectations.
    \benchmarkname{} tests each agent on progressively larger networks by increasing~$k$ and 
    identifies the maximum $k^*$ that satisfies both constraints.

\section{NetOps Agents}
\label{sec:agent}

\begin{figure}[t]
\centering
\includegraphics[width=\textwidth]{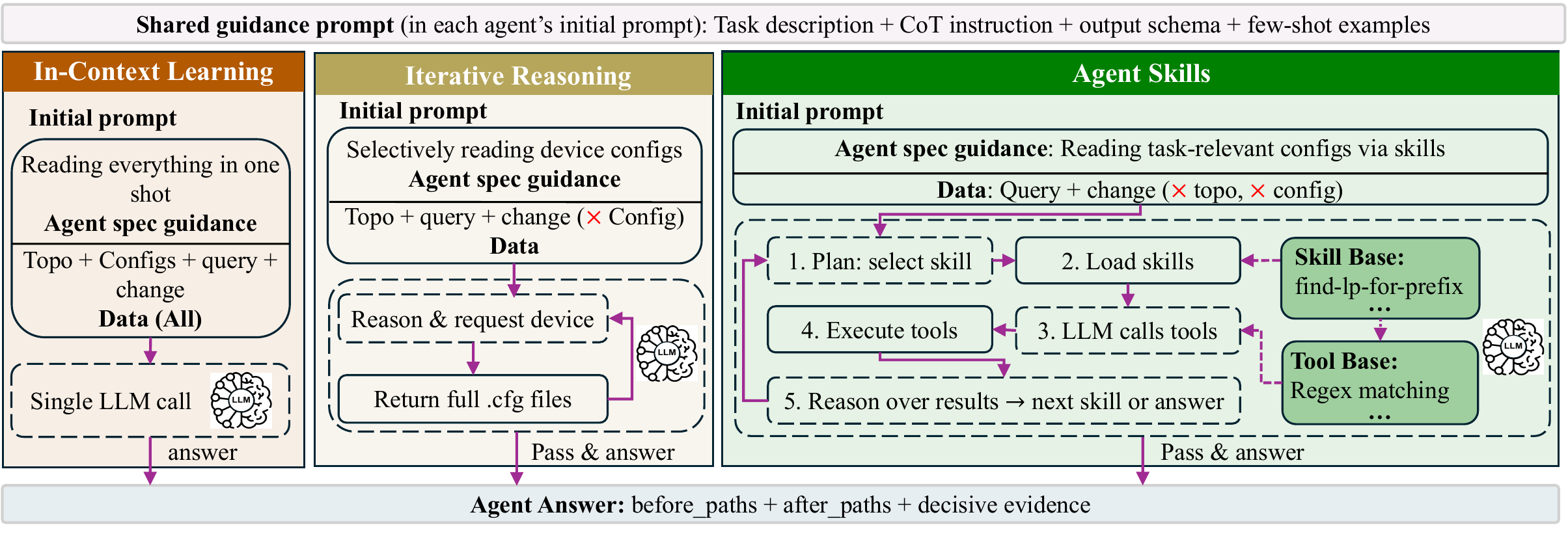}
\vspace{-15pt}
\caption{NetOps agents for path analysis in different agentic paradigms.}
\label{fig:workflow}
\vspace{-5pt}
\end{figure}

We develop three NetOps agents with different paradigms,
    {\it \methodnaive{}}, {\it \methodmulti{}}, 
    and {\it \methodagent{}}, as shown in  Figure~\ref{fig:workflow}.
The key insight is that the information needed to analyze routing paths is small in volume 
    but sparsely scattered across lengthy device configurations, 
    and retrieving it requires a long reasoning chain across protocols 
    (e.g., tracing a prefix list to the matching route-map and the LP statement, then reconciling with co-existing static routes).

These agents share a common \emph{guidance prompt} (\Cref{app:shared}) that specifies the path-analysis task following production operators' practice,
    including (1) input/output schema,
    (2) chain-of-thought (CoT) rules that instruct LLMs to trace paths, 
        compare routing attributes, and check reachability,
    and (3) few-shot demonstrations on small networks that ground expected outputs.
Each agentic method also has its own guidance (Figure~\ref{fig:workflow}) describing
    its evidence-access interface and retrieval workflow.

{\bf \methodnaive{}: Reading everything.}
The agent organizes the full network topology, all device configurations, the network change
    in one prompt (Figure~\ref{fig:workflow}).
The LLM then follows the shared CoT guidance to produce the paths.
We find that the \methodnaive{} agent is hard to scale, because the volume of all device configurations 
    greatly exceeds the context length 
    of existing LLMs, e.g., feeding GPT-5.4 all the device configurations of a single production datacenter
    already exceeds 1M tokens with only 300+ routers, which are far below the 10K-router scale of today's intra-datacenter networks,
    let alone inter-datacenter networks.

\textbf{\methodmulti{}: Selectively reading device configurations.}
The agent sets an initial prompt that includes the network topology and the network change,
    without any device configuration.
In each round, the agent selects which routers' configurations to inspect; 
    the system returns the complete \texttt{.cfg} file for each requested device.
The model analyzes the returned configuration, plans the next request, and iterates until it produces a final answer 
    or reaches a round limit.
In Figure~\ref{fig:path-analysis-example}, the agent may first request Aggr-1's 
    configuration to examine the LP change, then iteratively request configurations of neighboring devices 
    (e.g., Core-1 and -2) to verify whether the change redirects the path. Such analysis continues along 
    the new path, e.g., requesting FW and Peer-B when analyzing Border-2.
Although \methodmulti{} avoids loading all configurations at once, 
    each requested file is still raw and largely irrelevant 
    (the decisive evidence occupies only a few lines), 
    so the cumulative irrelevant input grows with network scale 
    and limits the agent's potential. 

\textbf{\methodagent{}: Reading task-relevant configurations via skills.}
The agent sets an initial prompt that includes only the network change, 
    without network topology or device configurations.
Instead of reading raw configuration files, the agent follows a \emph{plan-invoke-reason} loop 
    over routing-specific \emph{skills}: 
    it plans which skills to call, reasons over the returned information, 
    and either invokes additional skills or outputs the final answer. 
Each skill is a reusable, multi-step document that guides the LLM through a focused subtask, 
    optionally invoking tools~\cite{skills}, e.g., regular-expression search, 
    topology lookup, and configuration extraction.
Given a high-level objective like determining the LP for 
    destination prefix 10.2.0.0/16 on Aggr-1 in Figure~\ref{fig:path-analysis-example}, 
    a skill uses tools to extract the relevant configuration chain (e.g., prefix list $\to$ route map $\to$ LP) 
    and returns compact task-relevant statements or small configuration snippets (e.g., those in Figure~\ref{fig:path-analysis-example}). 
We developed 14~skills in three categories (see \Cref{app:agent-skills}): 
    seven for configuration extraction 
    (e.g., extracting prefix-specific LP from a route-map chain, checking interface reachability), 
    three for topology extraction (e.g., identifying candidate paths and enumerating neighbors), 
    and four verification skills (e.g., determining route selection, 
    verifying BGP session correctness, checking physical link and path reachability).
In the example of Figure~\ref{fig:path-analysis-example}, the agent 
    first invokes a \texttt{find\_candidate\_paths} skill 
    to enumerate the candidate paths (e.g., the blue and red paths), 
    then calls the \texttt{find\_lp\_for\_prefix} skill to 
    trace Aggr-1's LP for prefix 10.2.0.0/16 through the configuration chain (prefix list $\to$ route map 
    \texttt{LP\_Dst\_TO\_Core2} $\to$ \texttt{set local-preference~200}).
    Similarly, when analyzing Border-2, the agent invokes skills to trace and retrieve the static route and BGP MED evidence.
The \methodagent{} prompt also includes an accumulated set of \emph{don't} rules 
    that encode recurring failure anti-patterns as lightweight self-correction guidance (see \Cref{app:dont-rules}).

\section{Results}
\label{sec:results}

\subsection{Setup}

We evaluate three agent designs with six LLMs on \benchmarkname{}: 
  GPT-5.4 (2026-03-05), 
  GPT-4o (2024-11-20), 
  OpenAI o3 (2025-04-16)~\cite{GPT}, 
  DeepSeek-V4-Pro~\cite{deepseekv4pro}, 
  Qwen-3.5-397B-A17B~\cite{qwen35}, 
  and Llama-3.3-70B-Instruct~\cite{Llama33}.
Context windows are 1M tokens for GPT-5.4 and DeepSeek-V4-Pro, 
  262K for Qwen-3.5, 200K for OpenAI o3, and 128K for the other LLMs.
We set a one-hour time limit, matching the typical frequency of configuration changes in production (Table~\ref{tab:Configupdate}).
All experiments, including tool calls, run on a server with dual Intel Xeon Gold 6338 64-core processors 
  and 1\,TB RAM running Ubuntu 20.04.

{\bf Task Set.}
For intra-datacenter networks, we generate $100$ instances per configuration-change type 
  (See Table~\ref{tab:advanced-configs})---$600$ per network size $k$---for 
  the three agents with GPT-5.4 and GPT-4o,
  and $30$ per type ($180$ per $k$) for other models due to rate limits.
We evaluate $k$ up to $k{=}200$ (50K routers), where GPT-5.4 hits its limit.
An agent reaches its scalability limit at network size $k$ 
  if 25\% of tasks at that size exceed the context window or cannot finish in an hour.
For inter-datacenter and compound changes, we use GPT-5.4 and GPT-4o 
  with 100 instances per change type across the same scales.

\subsection{Benchmark Results}
\label{sec:overall}

{\bf Promises.}
On intra-datacenter single-change analysis, \methodagent{} with GPT-5.4 can scale to 50K routers ($k{=}200$) 
  at 99.5\% accuracy using less than 120K tokens, 
  while Batfish, the state-of-the-art symbolic analysis, runs out of memory beyond 8K routers (Figure~\ref{fig:batfish-runtime}).
The results show the promising potential of agentic approaches to scale routing-path analysis.

\begin{figure}[t]
\centering
\includegraphics[width=.9\linewidth]{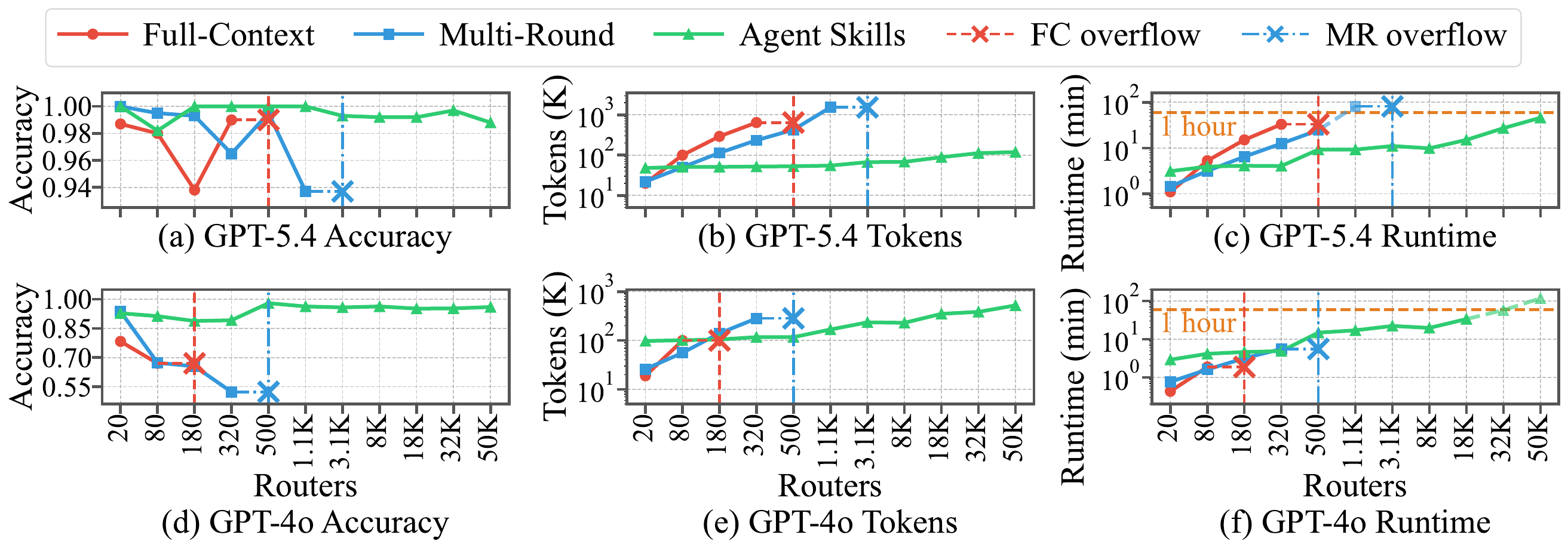}
\vspace{-5pt}
\caption{Benchmark results of three agents with GPT-5.4 (top) and GPT-4o (bottom) 
  on intra-datacenter networks, 
  averaged over all types of individual changes. See~\Cref{app:all-results} for full results.}
\label{fig:three-methods}
\vspace{-12pt}
\end{figure}

Figure~\ref{fig:three-methods} shows that, without carefully engineered skills, 
  neither raw model nor simple planning scales well. 
Specifically, \methodnaive{} overflows the context window at 500 and 180 routers with GPT-5.4 and GPT-4o respectively, 
  and \methodmulti{} only pushes the scale to 3.1K and 500 routers, respectively;
  moreover, the analysis accuracy significantly decreases on the scalability boundary.
On the other hand,
  \methodagent{} scales to 50K routers for both LLMs at near-100\% accuracy, 
  with the most stable token and runtime curves across agent designs: 
  tokens stay below 119K (GPT-5.4) and 360K (GPT-4o).
At scale, running time grows faster than token usage, as the bottleneck shifts 
  from LLM planning to tool execution (e.g., searching topology paths)
  whose cost grows with network size rather than context.
\methodagent{} with other LLMs likewise hit the time limit before exhausting 
  their context windows (Table~\ref{tab:main-results}).

As shown in Table~\ref{tab:main-results}, the capabilities of LLMs matter. 
In terms of scalability, we argue that the deciding factor is the cost.
GPT-5.4 is the most cost-efficient---it uses the fewest LLM rounds, while invoking the most tools 
  per round and reaches the same accuracy with ${\sim}3{\times}$ 
  fewer tokens than thinking-oriented models such as OpenAI o3, DeepSeek-V4-Pro, and Qwen-3.5 
  (which prefer internal reasoning over tool calls in a round).
Weaker models like GPT-4o and Llama-3.3 lack confidence in skill-based analysis and need more reasoning rounds 
  (e.g., GPT-4o uses $2{\times}$ of LLM rounds of GPT-5.4), 
  inflating both tokens and time (Figure~\ref{fig:three-methods}).
We distill these observations into a unified principle: \emph{explore more; digest less}---aggressively 
  invoke skills with tools, instead of reading raw configurations or over-thinking internally.
\begin{table}[t]
\centering
\caption{\methodagent{} results 
  across LLMs on intra-datacenter networks, 
  averaged over scales and 
  types of individual changes within the time/context limit. 
  $R_{\mathrm{LLM}}$/$R_{\mathrm{Tool}}$: average rounds of LLM calls (each outputs a JSON tool-invocation plan) and tool invocations (possibly multiple per LLM round) per task.
  $^\dagger$: bottleneck at the next scale (time limit or context window overflow).}
\label{tab:main-results}
\vspace{3pt}
\footnotesize
\setlength{\tabcolsep}{4pt}
\begin{tabular}{@{}llrrrrrrr@{}}
\toprule
Rank & Model & $k^{*}$ & Accuracy & Tokens & $t_{\mathrm{LLM}}$ & $t_{\mathrm{Runtime}}$ & $R_{\mathrm{LLM}}$ & $R_{\mathrm{Tool}}$ \\
\midrule
1 & GPT-5.4         & 200 (50K)             & \accH{99.5\%} & \tokVL{ 70K} & \tlL{ 212 s} & \rtL{ 778 s} & \rndL{ 7.8} & \rtoH{29.5} \\
2 & Qwen-3.5-397B   & 120 (18K)$^\dagger$   & \accH{99.3\%} & \tokM{231K}  & \tlL{ 351 s} & \rtM{1101 s} & \rndM{18.9} & \rtoM{22.6} \\
3 & DeepSeek-V4-Pro & 160 (32K)$^\dagger$   & \accH{99.1\%} & \tokM{178K}  & \tlM{ 543 s} & \rtM{1331 s} & \rndM{15.8} & \rtoH{29.2} \\
4 & OpenAI o3       & 160 (32K)$^\dagger$   & \accHM{96.8\%} & \tokM{208K}  & \tlL{ 323 s} & \rtM{1247 s} & \rndM{18.0} & \rtoL{12.0} \\
5 & GPT-4o          & 120 (18K)$^\dagger$   & \accM{93.8\%} & \tokM{169K}  & \tlVL{ 172 s} & \rtL{ 830 s} & \rndM{15.7} & \rtoL{15.9} \\
6 & Llama-3.3-70B   & 160 (32K)$^\dagger$   & \accLM{73.3\%} & \tokH{408K} & \tlH{ 752 s} & \rtH{1910 s} & \rndH{23.5} & \rtoM{24.3} \\
\bottomrule
\end{tabular}
\vspace{-10pt}
\end{table}

{\bf Boundary.}
\label{sec:compound}
\benchmarkname{} also shows the boundary of the evaluated agents. 
Tables~\ref{tab:complex-crossdc} and~\ref{tab:complex-compound} show 
the results of \methodagent{} with GPT-5.4 under inter-datacenter and compound-change scenarios.
The analysis scales to 100K+ routers in inter-datacenter cases with 70+\% accuracy,
  but starts to decrease beyond that.
The analysis errors follow two patterns: (1) in inter-datacenter cases, lazy reasoning over long paths (8 hops vs.\ 5 in intra-datacenter) 
  causes the agent to skip per-router analysis and fall back to default next hops (e.g., lowest router-ID); 
  (2) in compound changes, the agent struggles when multiple routing protocols affect the 
  same flow on different routers.
We expect more capable models to push this boundary further. 

\begin{table}[H]
\vspace{-7.5pt}
\centering
\small
\begin{minipage}[t]{0.48\textwidth}
\centering
\captionof{table}{\methodagent{} (GPT-5.4) results on inter-datacenter networks at selected scales.}
\label{tab:complex-crossdc}
\resizebox{\linewidth}{!}{%
\begin{tabular}{@{}llrrrrr@{}}
\toprule
$k$ & Routers & Accuracy & Tokens & $t_{\mathrm{LLM}}$ & $t_{\mathrm{Runtime}}$ & $R_{\mathrm{LLM}}$ \\
\midrule
  4 &     48 &\accHM{96.2\%}&\tokL{91K}&\tlL{276s}&\rtVL{329s}&\rndL{7.6}\\
 16 &    672 &\accM{90.0\%}&\tokL{100K}&\tlL{305s}&\rtVL{418s}&\rndL{8.0}\\
 50 &  6,350 &\accLM{83.7\%}&\tokL{99K}&\tlL{302s}&\rtL{921s}&\rndL{7.9}\\
200 & 100,400 &\accLM{73.5\%}&\tokL{127K}&\tlL{387s}&\rtH{3012s}&\rndL{9.2}\\
\bottomrule
\end{tabular}%
}
\end{minipage}\hfill
\begin{minipage}[t]{0.48\textwidth}
\centering
\captionof{table}{\methodagent{} (GPT-5.4) results on compound changes at selected scales.}
\label{tab:complex-compound}
\resizebox{\linewidth}{!}{%
\begin{tabular}{@{}llrrrrr@{}}
\toprule
$k$ & Routers & Accuracy & Tokens & $t_{\mathrm{LLM}}$ & $t_{\mathrm{Runtime}}$ & $R_{\mathrm{LLM}}$ \\
\midrule
  4 &     20 &\accH{99.5\%}&\tokVL{48K}&\tlVL{147s}&\rtVL{178s}&\rndVL{6.4}\\
 16 &    320 &\accHM{96.2\%}&\tokVL{58K}&\tlVL{177s}&\rtVL{245s}&\rndVL{6.4}\\
 50 &  3,125 &\accM{80.5\%}&\tokVL{61K}&\tlVL{187s}&\rtL{571s}&\rndVL{6.5}\\
200 & 50,000 &\accLM{79.0\%}&\tokVL{58K}&\tlVL{176s}&\rtH{2100s}&\rndVL{6.3}\\
\bottomrule
\end{tabular}%
}
\end{minipage}
\vspace{-3pt}
\end{table}

{\bf On Change Types.}
\label{sec:changetype}
Table~\ref{tab:change-breakdown} shows the accuracy results of
  \methodagent{} with GPT-5.4 for each change type.
For individual changes, analysis errors only occur on MED, 
  the only type where a network flow cannot decide its next hop directly. 
It receives a network flow from a neighbor and must compete on MED values across all routers connected to that neighbor, 
  whereas LP and static route either compete directly among a router's own neighbors 
  or directly specify the next step. 
  This indirection becomes harder as the number of routers and links grows.
For compound changes, only LP+Shut keeps high accuracy since (physical) shutdown effectively 
  reduces the task to an individual routing-protocol change. 
The other combinations involve multiple protocols controlling the same flow, 
  and the agent gets confused when competing protocols span different routers, 
  often considering only one change and ignoring the other. 
This confusion worsens with scale where router count increases. 
\begin{table}[t]
\centering
\caption{\methodagent{} (GPT-5.4) accuracy by change type at selected scales.}
\label{tab:change-breakdown}
\scriptsize
\resizebox{\textwidth}{!}{%
\begin{tabular}{@{}llrrrrrrrrrr@{}}
\toprule
$k$ & Routers & LP & MED & Static & Shut & BGP Net & Pfx Agg & LP+Net & LP+Shut & LP+St+Shut & LP+St+Net+Shut \\
\midrule
  4 &     20 & \accH{100\%} & \accH{100\%} & \accH{100\%} & \accH{100\%} & \accH{100\%} & \accH{100\%} & \accH{99\%}  & \accH{99\%}  & \accH{100\%} & \accH{100\%} \\
 16 &    320 & \accH{100\%} & \accH{100\%} & \accH{100\%} & \accH{100\%} & \accH{100\%} & \accH{100\%} & \accM{95\%}  & \accH{100\%} & \accM{95\%}  & \accM{95\%}  \\
 50 &  3,125 & \accH{100\%} & \accHM{96\%}  & \accH{100\%} & \accH{100\%} & \accH{100\%} & \accH{100\%} & \accLM{76\%} & \accH{100\%} & \accLM{73\%} & \accLM{73\%} \\
200 & 50,000 & \accH{100\%} & \accM{93\%}  & \accH{100\%} & \accH{100\%} & \accH{100\%} & \accH{100\%} & \accLM{73\%} & \accH{100\%} & \accLM{71\%} & \accLM{72\%} \\
\bottomrule
\end{tabular}%
}
\vspace{-15pt}
\end{table}

\subsection{Evaluation on a Real-World Hyperscale Network}
\label{sec:realnet}

To check the fidelity of \benchmarkname{},
    we evaluate \methodagent{} with GPT-5.4 on a real-world 
    hyperscale production network.
The network includes two datacenters with $10K$+ routers each, connected by $100$+ routers 
(the same network reported in Figure~\ref{fig:config-cdf}), 
    with ground truth derived from production routing tables.
We created 100 intra- and 100 inter-datacenter analysis queries.
\methodagent{} achieves 92\% accuracy on intra-datacenter analysis
    and 73\% on inter-datacenter analysis (see Table~\ref{tab:realnet-gpt54}).
The results are consistent with those measured by \benchmarkname{} 
    (90+\% for intra- and 70+\% for inter-datacenter analysis).

We find higher token consumption and LLM time on the real-world network than on \benchmarkname{}, because real-world configurations mix routing with other 
    complexities (traffic bandwidth allocation, device login authentication, etc.).
\benchmarkname{} currently only includes routing-related components.
Our future work includes supporting other kinds of configurations in \benchmarkname{}.
\section{Related Work}

Routing-path analysis, which essentially reasons about how target networks 
    route packets, is a keystone of NetOps.
The scalability and accuracy of path analysis 
    determine network reliability and security.
\benchmarkname{} is the first benchmark of this fundamental task.
{\bf NetOps Agents.}
Existing NetOps agents mainly target two tasks:
(1) failure diagnosis for identifying 
    root causes of network failures 
    from observability data~\cite{netassistant,aidai}, and
(2) configuration generation 
    that synthesizes new configurations to realize high-level intent~\cite{cegs,netconfeval,confucius,NetLLM,intent-extract,protocol-understand,topo-understand,topo-understand2}.
No prior agents study path analysis.
Path analysis is complementary to failure diagnosis---it is commonly used in prevention 
    of failures and also as a key primitive for diagnosis.
In fact, with the rise of AI-generated device configurations,
    accurate and scalable path analysis has become more important than ever
    to understand these configurations and prevent configuration-induced failures.

\noindent\textbf{NetOps Benchmarks.}
Existing NetOps benchmarks such as NetArena~\cite{netarena}
    and NIKA~\cite{NIKA}
    primarily focus on failure diagnosis or troubleshooting,
    with device misconfigurations being major causes of network failures~\cite{fb-outage-2021,azure-outage-2023,google-outage-2020}.
\benchmarkname{} instead focuses on routing-path analysis as an essential primitive
    for preventing configuration-induced network failures 
    and for providing basic utilities for troubleshooting.
We believe that the idea of automatically generating realistic networks in \benchmarkname{}
    can benefit other kinds of NetOps-related benchmarks.

\section{Concluding Remarks}

We explore an intriguing question---whether one could use an agentic approach
    to scale routing-path analysis effectively for hyperscale networks where traditional symbolic tools fail (while maintaining high accuracy).
While the requirements of path analysis bring significant challenges to frontier LLMs,
    we find that, with the principle of ``explore more; digest less'',
    carefully engineered \methodagent{} yield promising results on \benchmarkname{} (which is also 
    validated in production networks).
We will use \benchmarkname{} to continuously push the boundary of
    agentic AI technologies for NetOps tasks towards reliable, secure, and autonomous hyperscale network infrastructures.
\renewcommand{\acksection}{\section*{Acknowledgments}}
\begin{ack}
Zhixiong Niu and Hong Xu are co-corresponding authors.
\end{ack}

{
\small
\bibliographystyle{plainnat}
\bibliography{reference}

@misc{GPT,
  author = {OpenAI},
    year = {2025},
	howpublished = {\url{https://openai.com/chatgpt/overview/}},
	title = {{OpenAI ChatGPT}}}

@article{topo-understand2,
      title={{Large Language Model (LLM)-enabled Graphs in Dynamic Networking}},
      author={Geng Sun and Yixian Wang and Dusit Niyato and Jiacheng Wang and Xinying Wang and H. Vincent Poor and Khaled B. Letaief},
      year={2024},
      journal = {arXiv preprint arXiv:2407.20840},
}

@inproceedings{s2, 
  author = {Wang, Dan and Zhang, Peng and Sun, Wenbing and Li, Wenkai and Feng, Xing and Li, Hao and Chen, Jiawei and Jiang, Weirong and Tang, Yongping}, 
  title = {{S2: A Distributed Configuration Verifier for Hyper-Scale Networks}}, 
  year = {2025}, 
  booktitle = {Proceedings of the ACM Conference on Special Interest Group on Data Communication (SIGCOMM)}
}

@inproceedings{chainofthought,
  title={{Chain-of-Thought Prompting Elicits Reasoning in Large Language Models}},
  author={Wei, Jason and Wang, Xuezhi and Schuurmans, Dale and Bosma, Maarten and Xia, Fei and Chi, Ed and Le, Quoc V and Zhou, Denny and others},
  booktitle={Advances in Neural Information Processing Systems (NeurIPS)},
  year={2022}
}

@misc{acl-cisco,
  author = {Cisco},
  year = {2025},
	howpublished = {\url{https://www.cisco.com/c/en/us/support/docs/security/ios-firewall/23602-confaccesslists.html}},
	title = {{Cisco ACL}}}

@misc{bgp-cisco,
  author = {Cisco},
    year = {2025},
	howpublished = {\url{https://www.cisco.com/c/en/us/support/docs/ip/border-gateway-protocol-bgp/13751-23.html}},
	title = {{Cisco BGP}}}

@inproceedings{Minesweeper,
  title = {A {{General Approach}} to {{Network Configuration Verification}}},
  booktitle = {Proceedings of the ACM Conference on Special Interest Group on Data Communication (SIGCOMM)},
  author = {Beckett, Ryan and Gupta, Aarti and Mahajan, Ratul and Walker, David},
  year = {2017}
}

@inproceedings{ProbNetKAT,
  title = {Cantor {{Meets Scott}}: {{Semantic Foundations}} for {{Probabilistic Networks}}},
  booktitle = {Proceedings of the ACM SIGPLAN Symposium on Principles of Programming Languages (POPL)},
  author = {Smolka, Steffen and Kumar, Praveen and Foster, Nate and Kozen, Dexter and Silva, Alexandra},
  year = {2017}
}

@inproceedings{QARC,
  title = {Detecting {{Network Load Violations}} for {{Distributed Control Planes}}},
  booktitle = {Proceedings of the ACM SIGPLAN Conference on Programming Language Design and Implementation (PLDI)},
  author = {Subramanian, Kausik and Abhashkumar, Anubhavnidhi and D'Antoni, Loris and Akella, Aditya},
  year = {2020}
}

@inproceedings{ARC,
  title = {Fast {{Control Plane Analysis Using}} an {{Abstract Representation}}},
  booktitle = {Proceedings of the ACM Conference on Special Interest Group on Data Communication (SIGCOMM)},
  author = {{Gember-Jacobson}, Aaron and Viswanathan, Raajay and Akella, Aditya and Mahajan, Ratul},
  year = {2016}
}

@inproceedings{NetDice,
  title = {Probabilistic {{Verification}} of {{Network Configurations}}},
  booktitle = {Proceedings of the ACM Conference on Special Interest Group on Data Communication (SIGCOMM)},
  author = {Steffen, Samuel and Gehr, Timon and Tsankov, Petar and Vanbever, Laurent and Vechev, Martin},
  year = {2020}
}

@inproceedings{rela,
  title = {Relational {{Network Verification}}},
  booktitle = {Proceedings of the ACM Conference on Special Interest Group on Data Communication (SIGCOMM)},
  author = {Xu, Xieyang and Yuan, Yifei and Kincaid, Zachary and Krishnamurthy, Arvind and Mahajan, Ratul and Walker, David and Zhai, Ennan},
  year = {2024}
}

@misc{cisco-interface-shutdown,
  author = {Cisco},
  year = {2025},
  title={{Cisco Interface Shutdown}},
  howpublished = {\url{https://www.cisco.com/E-Learning/bulk/public/tac/cim/cib/using_cisco_ios_software/cmdrefs/shutdown.htm}}
}

@misc{cisco-local-preference,
  author = {Cisco},
  title={{Cisco BGP Local Preference}},
  howpublished = {\url{https://www.cisco.com/c/en/us/td/docs/ios-xml/ios/iproute_bgp/configuration/15-mt/irg-15-mt-book/irg-external-sp.html#GUID-CD3F70AE-C92B-4A6D-AE85-460B6BCDBB17}},
  year = {2014}
}

@misc{cisco-bgp-network,
  author = {Cisco},
  title={{Cisco BGP Network Advertisement}},
  howpublished = {\url{https://www.cisco.com/c/en/us/support/docs/ip/border-gateway-protocol-bgp/16137-cond-adv.html}},
  year = {2025}
}

@misc{cisco-med,
  author = {Cisco},
  title={{Cisco BGP Multi-Exit Discriminator (MED)}},
  howpublished = {\url{https://www.cisco.com/c/en/us/support/docs/ip/border-gateway-protocol-bgp/13759-37.html}},
  year = {2024}
}

@misc{cisco-aggregate-route,
  author = {Cisco},
  title={{Cisco BGP Route Aggregation}},
  howpublished = {\url{https://www.cisco.com/c/en/us/support/docs/ip/border-gateway-protocol-bgp/5441-aggregation.html}},
  year = {2024}
}

@misc{junos-bgp-local-preference,
  author = {Juniper Networks},
  title={{Juniper BGP Local Preference}},
  howpublished = {\url{https://www.juniper.net/documentation/us/en/software/junos/bgp/topics/topic-map/local-preference.html}},
  year = {2025}
}

@misc{junos-bgp-med,
  author = {Juniper Networks},
  title={{Juniper BGP MED Attribute}},
  howpublished = {\url{https://www.juniper.net/documentation/us/en/software/junos/bgp/topics/topic-map/med-attribute.html}},
  year = {2025}
}

@misc{junos-bgp-network,
  author = {Juniper Networks},
  title={{Juniper BGP Network Advertisement}},
  howpublished = {\url{https://www.juniper.net/documentation/us/en/software/junos/routing-policy/bgp/topics/example/bgp-advertise-peer-as.html}},
  year = {2025}
}

@misc{junos-static-routes,
  author = {Juniper Networks},
  title={{Juniper Static Routes}},
  howpublished = {\url{https://www.juniper.net/documentation/us/en/software/junos/static-routing/topics/topic-map/config_static-routes.html}},
  year = {2025}
}

@misc{junos-aggregate-route,
  author = {Juniper Networks},
  title={{Juniper Route Aggregation}},
  howpublished = {\url{https://www.juniper.net/documentation/us/en/software/junos/static-routing/topics/topic-map/config-route-aggregation.html}},
  year = {2025}
}

@misc{junos-interface-shutdown,
  author = {Juniper Networks},
  title={{Juniper Interface Shutdown}},
  howpublished = {\url{https://www.juniper.net/documentation/us/en/software/junos/cli-reference/topics/ref/statement/interface-shutdown-action-edit-switch-options.html}},
  year = {2025}
}

@article{netconfeval,
  title={{NetConfEval: Can LLMs Facilitate Network Configuration?}},
  author={Wang, Changjie and Scazzariello, Mariano and Farshin, Alireza and Ferlin, Simone and Kosti{\'c}, Dejan and Chiesa, Marco},
  journal={Proceedings of the ACM on Networking},
  volume={2},
  number={CoNEXT2},
  pages={1--25},
  month={June},
  year={2024}
}

@article{relationalnetkat,
author = {Xu, Han and Kincaid, Zachary and Mahajan, Ratul and Walker, David},
title = {{Network Change Validation with Relational NetKAT}},
year = {2026},
volume = {10},
journal = {Proceedings of the ACM on Programming Languages},
month = jan,
number = {14},
pages = {384--412}
}

@misc{cisco-static,
  author = {Cisco},
  year = {2011},
	howpublished = {\url{https://www.cisco.com/c/en/us/td/docs/switches/datacenter/nexus3000/sw/unicast/503_u1_2/nexus3000_unicast_config_gd_503_u1_2/l3_route.html}},
	title = {{Cisco Static Route}}}

@inproceedings{swan,
	author = {Hong, Chi-Yao and Kandula, Srikanth and Mahajan, Ratul and Zhang, Ming and Gill, Vijay and Nanduri, Mohan and Wattenhofer, Roger},
	booktitle = {Proceedings of the ACM Conference on Special Interest Group on Data Communication (SIGCOMM)},
	title = {{Achieving High Utilization with Software-Driven WAN}},
	year = {2013}}

@misc{azure,
  author = {Microsoft},
  howpublished = {\url{https://azure.microsoft.com/en-us/explore/global-infrastructure}},
  title = {{Azure Global DC Network}},
  year = {2025}
}

@inproceedings{fattree1,
author = {Kandula, Srikanth and Sengupta, Sudipta and Greenberg, Albert and Patel, Parveen and Chaiken, Ronnie},
title = {{The Nature of Data Center Traffic: Measurements \& Analysis}},
year = {2009},
booktitle = {Proceedings of the ACM Internet Measurement Conference (IMC)}
}

@inproceedings{fattree2,
author = {Al-Fares, Mohammad and Loukissas, Alexander and Vahdat, Amin},
title = {{A Scalable, Commodity Data Center Network Architecture}},
year = {2008},
booktitle = {Proceedings of the ACM Conference on Special Interest Group on Data Communication (SIGCOMM)}
}

@inproceedings{batfish,
author = {Fogel, Ari and Fung, Stanley and Pedrosa, Luis and Walraed-Sullivan, Meg and Govindan, Ramesh and Mahajan, Ratul and Millstein, Todd},
title = {{A General Approach to Network Configuration Analysis}},
year = {2015},
booktitle = {Proceedings of the USENIX Symposium on Networked Systems Design and Implementation (NSDI)}
}

@inproceedings{batfish2,
author = {Brown, Matt and Fogel, Ari and Halperin, Daniel and Heorhiadi, Victor and Mahajan, Ratul and Millstein, Todd},
title = {{Lessons from the Evolution of the Batfish Configuration Analysis Tool}},
year = {2023},
booktitle = {Proceedings of the ACM Conference on Special Interest Group on Data Communication (SIGCOMM)}
}

@inproceedings{cosynth,
author = {Mondal, Rajdeep and Tang, Alan and Beckett, Ryan and Millstein, Todd and Varghese, George},
title = {{What do LLMs Need to Synthesize Correct Router Configurations?}},
year = {2023},
booktitle = {Proceedings of the ACM Workshop on Hot Topics in Networks (HotNets)}
}

@inproceedings {cegs,
author = {Jianmin Liu and Li Chen and Dan Li and Yukai Miao},
title = {{CEGS: Configuration Example Generalizing Synthesizer}},
booktitle = {Proceedings of the USENIX Symposium on Networked Systems Design and Implementation (NSDI)},
year = {2025}
}

@inproceedings{netcomplete,
author = {Ahmed El-Hassany and Petar Tsankov and Laurent Vanbever and Martin Vechev},
title = {{NetComplete: Practical Network-Wide Configuration Synthesis with Autocompletion}},
booktitle = {Proceedings of the USENIX Symposium on Networked Systems Design and Implementation (NSDI)},
year = {2018}
}

@inproceedings {NDD,
author = {Zechun Li and Peng Zhang and Yichi Zhang and Hongkun Yang},
title = {{NDD: A Decision Diagram for Network Verification}},
booktitle = {Proceedings of the USENIX Symposium on Networked Systems Design and Implementation (NSDI)},
year = {2025}
}

@INPROCEEDINGS{topo-understand,
  author={Donadel, Denis and Marchiori, Francesco and Pajola, Luca and Conti, Mauro},
  booktitle={2024 IEEE 49th Conference on Local Computer Networks (LCN)}, 
  title={{Can LLMs Understand Computer Networks? Towards a Virtual System Administrator}}, 
  year={2024}}

@inproceedings{protocol-understand, 
    author = {Sharma, Prakhar and Yegneswaran, Vinod}, 
    booktitle = {Proceedings of the ACM Workshop on Hot Topics in Networks (HotNets)},
    title = {{PROSPER: Extracting Protocol Specifications Using Large Language Models}}, 
    year = {2023}, 
}

@INPROCEEDINGS{intent-extract,
  author={Manias, Dimitrios Michael and Chouman, Ali and Shami, Abdallah},
  booktitle={2024 20th International Conference on the Design of Reliable Communication Networks (DRCN)}, 
  title={{Towards Intent-Based Network Management: Large Language Models for Intent Extraction in 5G Core Networks}}, 
  year={2024}}

@misc{deepseekv4pro,
      title={{DeepSeek V4 Preview Release}},
      author={DeepSeek-AI},
      year={2026},
      howpublished={\url{https://api-docs.deepseek.com/news/news260424}},
}

@inproceedings{NetLLM,
   title={{NetLLM: Adapting Large Language Models for Networking}},
   booktitle={Proceedings of the ACM Conference on Special Interest Group on Data Communication (SIGCOMM)},
   author={Wu, Duo and Wang, Xianda and Qiao, Yaqi and Wang, Zhi and Jiang, Junchen and Cui, Shuguang and Wang, Fangxin},
   year={2024}}

@inproceedings {onewan,
author = {Umesh Krishnaswamy and Rachee Singh and Paul Mattes and Paul-Andre C Bissonnette and Nikolaj Bj{\o}rner and Zahira Nasrin and Sonal Kothari and Prabhakar Reddy and John Abeln and Srikanth Kandula and Himanshu Raj and Luis Irun-Briz and Jamie Gaudette and Erica Lan},
title = {{OneWAN is Better than Two: Unifying a Split WAN Architecture}},
booktitle = {Proceedings of the USENIX Symposium on Networked Systems Design and Implementation (NSDI)},
year = {2023}
}

@inproceedings{azure-dc,
author = {Jayaraman, Karthick and Bj\o{}rner, Nikolaj and Padhye, Jitu and Agrawal, Amar and Bhargava, Ashish and Bissonnette, Paul-Andre C and Foster, Shane and Helwer, Andrew and Kasten, Mark and Lee, Ivan and Namdhari, Anup and Niaz, Haseeb and Parkhi, Aniruddha and Pinnamraju, Hanukumar and Power, Adrian and Raje, Neha Milind and Sharma, Parag},
title = {{Validating datacenters at scale}},
booktitle = {Proceedings of the ACM Conference on Special Interest Group on Data Communication (SIGCOMM)},
year = {2019}
}

@inproceedings{confucius,
author = {Wang, Zhaodong and Lin, Samuel and Yan, Guanqing and Ghorbani, Soudeh and Yu, Minlan and Zhou, Jiawei and Hu, Nathan and Baruah, Lopa and Peters, Sam and Kamath, Srikanth and Yang, Jerry and Zhang, Ying},
title = {{Intent-Driven Network Management with Multi-Agent LLMs: The Confucius Framework}},
year = {2025},
booktitle = {Proceedings of the ACM Conference on Special Interest Group on Data Communication (SIGCOMM)}
}

@inproceedings {dna,
author = {Peng Zhang and Aaron Gember-Jacobson and Yueshang Zuo and Yuhao Huang and Xu Liu and Hao Li},
title = {{Differential Network Analysis}},
booktitle = {Proceedings of the USENIX Symposium on Networked Systems Design and Implementation (NSDI)},
year = {2022}
}

@inproceedings{netarena,
  title={{NetArena: Dynamic Benchmarks for AI Agents in Network Automation}},
  author={Yajie Zhou and Jiajun Ruan and Eric S. Wang and Sadjad Fouladi and Francis Y. Yan and Kevin Hsieh and Zaoxing Liu},
  booktitle={Proceedings of the International Conference on Learning Representations (ICLR)},
  year={2026}
}

@article{NIKA,
      title={{A Network Arena for Benchmarking AI Agents on Network Troubleshooting}},
      author={Zhihao Wang and Alessandro Cornacchia and Alessio Sacco and Franco Galante and Marco Canini and Dingde Jiang},
      year={2025},
      journal = {arXiv preprint arXiv:2512.16381},
}

@inproceedings{netassistant,
  title={{NetAssistant: Dialogue Based Network Diagnosis in Data Center Networks}},
  author={Haopei Wang and Anubhavnidhi Abhashkumar and Changyu Lin and Tianrong Zhang and Xiaoming Gu and Ning Ma and Chang Wu and Songlin Liu and Wei Zhou and Yongbin Dong and Weirong Jiang and Yi Wang},
  booktitle={Proceedings of the USENIX Symposium on Networked Systems Design and Implementation (NSDI)},
  year={2024}
}

@article{aidai,
      title={{AidAI: Automated Incident Diagnosis for AI Workloads in the Cloud}},
      author={Yitao Yang and Yangtao Deng and Yifan Xiong and Baochun Li and Hong Xu and Peng Cheng},
      year={2025},
      journal = {arXiv preprint arXiv:2506.01481},
}

@misc{skills,
  author = {Anthropic},
  year = {2025},
  title={{LLM Skills.}},
  howpublished = {\url{https://github.com/anthropics/skills}}
}

@misc{fb-outage-2021,
  title={{Understanding How Facebook Disappeared from the Internet}},
  author={Cloudflare},
  howpublished={\url{https://blog.cloudflare.com/october-2021-facebook-outage/}},
  year={2021}
}

@misc{google-outage-2020,
  title={{An Update on Sunday's Service Disruption}},
  howpublished={\url{https://cloud.google.com/blog/topics/inside-google-cloud/an-update-on-sundays-service-disruption}},
  year={2019},
  author={Google Cloud}
}

@misc{azure-outage-2023,
  title={{Post Incident Review (PIR) -- Azure Networking -- Global WAN issues (Tracking ID: VSG1-B90)}},
  howpublished={\url{https://azure.status.microsoft/en-us/status/history/?trackingId=VSG1-B90}},
  year={2023},
  author={Microsoft Azure}
}

@article{xumi,
      title={{Automating Conflict-Aware ACL Configurations with Natural Language Intents}},
      author={Ding, Wenlong and Li, Jianqiang and Niu, Zhixiong and Chen, Huangxun and Xiong, Yongqiang and Xu, Hong},
      year={2025},
      journal = {arXiv preprint arXiv:2508.17990},
}

@article{rag,
      title={{Retrieval-Augmented Generation for Knowledge-Intensive NLP Tasks}},
      author={Lewis, Patrick and Perez, Ethan and Piktus, Aleksandra and Petroni, Fabio and Karpukhin, Vladimir and Goyal, Naman and K{\"u}ttler, Heinrich and Lewis, Mike and Yih, Wen-tau and Rockt{\"a}schel, Tim and Riedel, Sebastian and Kiela, Douwe},
      year={2020},
      journal = {arXiv preprint arXiv:2005.11401},
}

@inproceedings{dragonfly,
  author    = {John Kim and William J. Dally and Steve Scott and Dennis Abts},
  title     = {Technology-Driven, Highly-Scalable Dragonfly Topology},
  booktitle = {Proceedings of the ACM/IEEE International Symposium on Computer Architecture (ISCA)},
  year      = {2008},
}

@misc{qwen35,
  title  = {{Qwen3.5-397B-A17B}},
  author = {Qwen Team},
  year   = {2026},
  howpublished = {\url{https://huggingface.co/Qwen/Qwen3.5-397B-A17B}},
}

@misc{Llama33,
  title  = {{Llama-3.3-70B-Instruct}},
  author = {Meta},
  year   = {2024},
  howpublished = {\url{https://huggingface.co/meta-llama/Llama-3.3-70B-Instruct}},
}

@inproceedings{jupiter,
  author = {Singh, Arjun and Ong, Joon and Agarwal, Amit and Anderson, Glen and Armistead, Ashby and Bannon, Roy and Boving, Seb and Desai, Gaurav and Felderman, Bob and Germano, Paulie and Kanagala, Anand and Provber, Jeff and Simmons, Jason and Tanda, Eiichi and Wanderer, Jim and H{\"o}lzle, Urs and Stuart, Stephen and Vahdat, Amin},
  title = {{Jupiter Rising: A Decade of Clos Topologies and Centralized Control in Google's Datacenter Network}},
  booktitle = {Proceedings of the ACM Conference on Special Interest Group on Data Communication (SIGCOMM)},
  year = {2015},
}

@misc{fb-fabric,
  author = {Andreyev, Alexey},
  title = {{Introducing Data Center Fabric, the Next-Generation Facebook Data Center Network}},
  year = {2014},
  howpublished = {\url{https://engineering.fb.com/2014/11/14/production-engineering/introducing-data-center-fabric-the-next-generation-facebook-data-center-network/}},
}

@inproceedings{HSA,
  title = {{Header Space Analysis: Static Checking for Networks}},
  booktitle = {Proceedings of the USENIX Symposium on Networked Systems Design and Implementation (NSDI)},
  author = {Kazemian, Peyman and Varghese, George and McKeown, Nick},
  year = {2012}
}
}

\newpage
\appendix

\providecommand{\toolname}[1]{\begingroup\def\_{\textunderscore\allowbreak}\texttt{#1}\endgroup}

\section{Additional Production Network Context}
\label{app:production-context}

\begin{table}[H]
\centering
\caption{Frequency of configuration changes in a global-scale production cloud network in the week of Feb.~26, 2025. Device metadata updates occur hourly; routing and filtering changes occur multiple times per day.}
\vspace{3pt}
\label{tab:Configupdate}
\renewcommand\arraystretch{1.25}
\resizebox{0.7\linewidth}{!}{
\begin{tabular}{l l cc}
\toprule
Config Type       & Remark                            & Percentage & Frequency             \\ \midrule
\begin{tabular}[c]{@{}l@{}}Device Metadata \\ \& Templates\end{tabular} &
  \begin{tabular}[c]{@{}l@{}}Prefix \& device addition/removal,\\ and device template setup.\end{tabular} &
  71.2\% &
  1.28 / hour \\
Routing Protocols & BGPs, IGPs, static routes, etc.   & 8.9\%      & 3.86 / day \\
Packet Filtering  & ACLs, firewalls, etc.             & 8.3\%      & 3.57 / day \\
Others            & NAT, device security config, etc. & 11.6\%     & 5.00 / day \\
\bottomrule
\end{tabular}}
\end{table}

\noindent\textbf{Configuration Change Frequency.} 
We measure how often configuration changes occur in a production network, since such changes can trigger {\it path analysis} to verify intended behavior.
As shown in Table~\ref{tab:Configupdate}, device-metadata changes, such as location and prefix updates, are the most frequent and occur hourly, prompting operators to check whether network paths are affected.
Routing-protocol and packet-filtering changes also occur multiple times per day, each potentially triggering path analysis to verify intended forwarding and reachability.
Overall, operators need path analysis at least hourly to keep pace with these changes while still leaving time to diagnose and resolve unexpected paths.

\section{Benchmark Generation Details}
\label{app:config-details}

All topology and configuration instances in \benchmarkname{} are produced by deterministic scripts parameterized by the fat-tree parameter~$k$ (even, $\ge 4$). 
This appendix details three layers of the generation pipeline: the physical topology and device/interface naming (\Cref{app:fattree-gen}), the connectivity configuration that ensures full eBGP reachability (\Cref{app:config-gen}), 
and the six types of flow-routing configuration changes used to construct benchmark tasks (\Cref{app:ad-config-gen}).

\subsection{Fat-Tree Topology Generation}
\label{app:fattree-gen}

\paragraph{Device naming.}
Each device is named by its role and position in the fat-tree hierarchy. Edge routers are \texttt{edge-p\{pod\}-e\{idx\}}, aggregation routers are \texttt{agg-p\{pod\}-a\{idx\}}, and core routers are \texttt{core-g\{group\}-c\{idx\}}, where $\text{pod} \in [0, k)$, $\text{idx} \in [0, k/2)$, and $\text{group} \in [0, k/2)$. Server hosts follow the pattern \texttt{server-p\{pod\}-e\{edge\}-h\{host\}}.
For inter-datacenter topologies, every device name is prefixed with a datacenter identifier \texttt{dc\{d\}-}, yielding names such as \texttt{dc0-edge-p0-e0} and \texttt{dc1-agg-p3-a1}. This prefix is the sole mechanism that distinguishes devices in the two datacenters---all index ranges remain identical, so \texttt{dc0-edge-p5-e2} and \texttt{dc1-edge-p5-e2} occupy the same structural position in their respective data centers.
Each datacenter additionally contains $k$~gateway routers named \texttt{dc\{d\}-gw\{g\}} ($g \in [0, k)$). Gateways connect to all $(k/2)^2$ cores in the local datacenter and to every gateway in the remote datacenter, providing inter-datacenter reachability.

\paragraph{Interface naming.}
We number each interface by the order in which it is added to the device, so the same $k$ always yields the same names. The order itself encodes role: lower-numbered interfaces face the layer below, higher-numbered ones face the layer above (e.g., on edge routers we add server-facing interfaces first, then uplinks to aggregation routers). Names may repeat across datacenters (e.g., both \texttt{dc0-gw0} and \texttt{dc1-gw0} have an interface \texttt{GigabitEthernet1/0/1}) without ambiguity because the device hostnames already carry the \texttt{dc\{d\}-} prefix.

\paragraph{Topology construction.}
The topology is built in three steps following the standard fat-tree wiring (Figure~\ref{fig:fattree}): (1)~each edge router connects to $k/2$ servers; (2)~within each pod, edge and aggregation routers form a complete bipartite graph; (3)~each aggregation router with index~$a$ connects to all $k/2$ core routers in core group~$a$. The output is a JSON file listing every link as a pair of \texttt{(hostname, interfaceName)} endpoints:

\begin{codebox}[Topology link entry (JSON)]
{"node1": {"hostname": "edge-p0-e0",
           "interfaceName": "GigabitEthernet1/0/3"},
 "node2": {"hostname": "agg-p0-a0",
           "interfaceName": "GigabitEthernet1/0/1"}}
\end{codebox}

For inter-datacenter, the script additionally wires each of the $k$~gateways per datacenter to all $(k/2)^2$ local cores, and establishes $k^2$ fully connected inter-datacenter gateway links, producing a single unified topology JSON.

\subsection{Network Connectivity Configuration}
\label{app:config-gen}
In \benchmarkname{}, we generate device configurations using a \textbf{Cisco~IOS-style} syntax~\cite{acl-cisco,bgp-cisco}, reflecting one of the most widely used router configuration formats.

\paragraph{Conflict-free IP addressing.}
All IP addresses are derived deterministically from each device's role and position, ensuring no prefix collisions across arbitrary~$k$. Table~\ref{tab:ip-scheme} summarizes the assignment scheme for intra-datacenter and inter-datacenter topologies.

\begin{table}[H]
\centering
\caption{IP address assignment by network layer (cf.\ Figure~\ref{fig:fattree}). DC1 and DC2 refer to the two datacenters in the inter-datacenter topology; intra-datacenter uses only DC1. Each /31 link entry lists the lower-tier side first, then the upper-tier side. Indices: $p$=pod, $e$=edge, $a$=agg, $c$=core, $g$=gateway, $\text{grp}$=core group, $\text{seq}$=index of a core-to-gateway link within a gateway.}
\label{tab:ip-scheme}
\scriptsize
\resizebox{.65\textwidth}{!}{%
\begin{tabular}{@{}ll@{}}
\toprule
Layer & Prefix assignment $\{$DC1, DC2$\}$ \\
\midrule
Server group       & \texttt{\{1,9\}.$p$.$e$.0/24}~~(gateway \texttt{.1}, hosts \texttt{.2+}) \\
Edge--Agg link     & \texttt{\{2,6\}.$p$.$a$.\{$2e$,$2e{+}1$\}/31} \\
Agg--Core link     & \texttt{\{3,7\}.$p$.$c$.\{$2a$,$2a{+}1$\}/31} \\
Core--GW link (inter-DC)      & \texttt{\{4,8\}.0.$g$.\{$2\,\text{seq}$,$2\,\text{seq}{+}1$\}/31} \\
GW--GW link (inter-DC) & \texttt{5.$g_0$.$g_1$.\{0,1\}/31} \\
Edge Loopback      & \texttt{10.$p$.\{0,2\}.$e$/32} \\
Agg Loopback       & \texttt{10.$p$.\{1,3\}.$a$/32} \\
Core Loopback      & \texttt{10.\{200,201\}.$\text{grp}$.$c$/32} \\
GW Loopback        & \texttt{10.\{250,251\}.0.$g$/32} \\
\bottomrule
\end{tabular}}
\end{table}

\noindent For intra-datacenter, server prefixes occupy the \texttt{1.x.x.x} space, edge--agg links use \texttt{2.x.x.x}, and agg--core links use \texttt{3.x.x.x}. Because the even/odd split within each /31 is determined by device indices, addresses never collide regardless of~$k$.
For inter-datacenter, datacenter~1 retains the \texttt{1/2/3/4} octets while datacenter~2 mirrors them at \texttt{9/6/7/8}, ensuring complete isolation between the two address domains. The inter-datacenter gateway links share the \texttt{5.x.x.x} space.

\noindent\textbf{Scaling beyond $k{=}256$ via IPv6.}
Each index field in Table~\ref{tab:ip-scheme} occupies one IPv4 octet (one byte), so $k$ caps at 256 (already over 80K routers). To scale further, we keep \emph{exactly the same layout} as Table~\ref{tab:ip-scheme} but widen every index field from one byte to one IPv6 hextet (two bytes), pushing the cap to $k{=}65{,}536$.

\noindent\textbf{VLAN for server-edge connectivity.}
Server endpoints use a different convention from inter-router links: instead of giving each link its own /31 IPs, all servers attached to the same edge router are placed in a virtual group (VLAN), share one /24 subnet, and use the edge router as their common gateway. This is a standard configuration on the server side. Each edge router uses a unique group ID ($1000{+}p\cdot k/2{+}e$), so no two groups in the network ever clash.

\begin{table}[H]
\centering
\caption{ASN assignment scheme. $d{\in}\{0,1\}$ indexes the datacenter so that the two datacenters fall into non-overlapping ranges; intra-datacenter uses $d{=}0$.}
\label{tab:asn-scheme}
\scriptsize
\resizebox{\textwidth}{!}{%
\begin{tabular}{@{}llp{7cm}@{}}
\toprule
Device Layer & Formula & Rationale \\
\midrule
Edge    & $100000 + d \cdot 10000 + p \cdot (k/2) + e$ & Every edge router gets its own ASN, so it never shares one with the aggregation routers it physically connects to. \\
Agg     & $20000 + d \cdot 1000 + p$ & All agg routers in the same pod share an ASN, but it is in a different range from edge ASNs, so every edge--agg link is between two different ASNs (i.e., eBGP). \\
Core    & $30000 + d$ & All core routers in a datacenter share an ASN, distinct from any agg ASN, so every agg--core link is again between two different ASNs. \\
Gateway & $40000 + d \cdot 1000 + g$ & Every gateway gets its own ASN, distinct from local cores and from every other gateway, so all core--gateway and gateway--gateway links are between two different ASNs. \\
\bottomrule
\end{tabular}}
\end{table}

\paragraph{eBGP session configuration.}
To ensure every inter-router link is an eBGP session (i.e., no two neighbors share an ASN), the ASN scheme assigns unique values per layer and scope. Table~\ref{tab:asn-scheme} lists the formulas.

Each router runs eBGP session on every point-to-point /31 interface. See the template below:

\begin{codebox}[eBGP configuration template (aggregation router)]
router bgp {ASN_AGG}
 bgp router-id {AGG_LOOPBACK}
 bgp log-neighbor-changes
 !
 address-family ipv4
  redistribute connected
  neighbor {EDGE_IP} remote-as {ASN_EDGE}
  neighbor {EDGE_IP} activate
  neighbor {CORE_IP} remote-as {ASN_CORE}
  neighbor {CORE_IP} activate
  neighbor {CORE_IP} route-map {RM_OUT} out
  aggregate-address {POD_PREFIX} {POD_MASK} as-set summary-only
  maximum-paths 8
 exit-address-family
!
\end{codebox}

\noindent The template parameters are filled as follows: \texttt{ASN\_AGG} is computed from Table~\ref{tab:asn-scheme}; \texttt{AGG\_LOOPBACK} and \texttt{EDGE\_IP}/\texttt{CORE\_IP} are derived from Table~\ref{tab:ip-scheme}; the route-map name encodes the pod and agg indices for traceability.

\begin{codebox}[eBGP configuration on agg-p0-a0 ($k$=4)]
router bgp 20000
 bgp router-id 10.0.1.0
 bgp log-neighbor-changes
 !
 address-family ipv4
  redistribute connected
  neighbor 2.0.0.0 remote-as 100000
  neighbor 2.0.0.0 activate
  neighbor 2.0.0.2 remote-as 100001
  neighbor 2.0.0.2 activate
  neighbor 3.0.0.1 remote-as 30000
  neighbor 3.0.0.1 activate
  neighbor 3.0.0.1 route-map RM_AGG_TO_CORE_0_0 out
  neighbor 3.0.1.1 remote-as 30000
  neighbor 3.0.1.1 activate
  neighbor 3.0.1.1 route-map RM_AGG_TO_CORE_0_0 out
  aggregate-address 1.0.0.0 255.255.0.0 as-set summary-only
  maximum-paths 8
 exit-address-family
!
\end{codebox}

\subsection{Flow Routing Configurations}
\label{app:ad-config-gen}

This subsection describes how the six configuration change types (Table~\ref{tab:advanced-configs-full}) are instantiated as benchmark tasks. Each type provides a \emph{template} parameterized by a source--destination flow $S(P_S) \to D(P_D)$, and shifts traffic to a different device by manipulating one specific routing attribute.

\begin{table}[t]
\centering
\caption{Six configuration types used to construct benchmark tasks (cf.\@ Table~\ref{tab:advanced-configs}).}
\label{tab:advanced-configs-full}
\small
\resizebox{\textwidth}{!}{%
\begin{tabular}{@{}p{3.6cm}p{3.4cm}p{8.5cm}@{}}
\toprule
Configuration Type & Category & Effect \\
\midrule
BGP Local Preference (LP)~\cite{cisco-local-preference,junos-bgp-local-preference} & Route preference & On a given router, a flow takes the next hop whose route-map sets the highest LP for that flow. \\
Multi-Exit Discriminator (MED)~\cite{cisco-med,junos-bgp-med} & Route preference & On a neighbor router, a flow chooses among the physical next hops connecting to it by comparing the MED values they advertise, and selects the one with the lowest MED. \\
Static route~\cite{cisco-static,junos-static-routes} & Route preference & Forces a flow onto a designated next hop, overriding any BGP-selected route. \\
BGP network advertisement~\cite{cisco-bgp-network,junos-bgp-network} & Route advertisement & Controls whether a router forwards a given flow via BGP; removing the \texttt{network} entry stops that router from forwarding the flow. \\
Prefix aggregation~\cite{cisco-aggregate-route,junos-aggregate-route} & Route advertisement & Summarizes multiple smaller flows into one aggregated flow forwarded via BGP; once a flow is covered by an aggregate, its own paths are suppressed and its forwarding follows the aggregated flow instead. \\
Interface shutdown~\cite{cisco-interface-shutdown,junos-interface-shutdown} & Topology availability & Physically disables an interface, so any flow traversing it can no longer be forwarded. \\
\bottomrule
\end{tabular}%
}
\end{table}

\noindent\textbf{BGP Local Preference (LP).}
\noindent The template is filled with: (i) \texttt{ASN}: the router's own ASN (Table~\ref{tab:asn-scheme}); (ii) \texttt{NEI\_IP}: the link IP of the chosen neighbor (Table~\ref{tab:ip-scheme}); (iii) \texttt{LP\_HIGH}: any value above the default LP. \textbf{\emph{The effect:}} among all of this router's neighbors, the one at \texttt{NEI\_IP} now has the highest LP for the queried flow, so the flow is shifted onto that neighbor.

\begin{codebox}[LP template]
route-map {RM_NAME} permit 10
 set local-preference {LP_HIGH}
route-map {RM_NAME} permit 20
!
router bgp {ASN}
 address-family ipv4
  neighbor {NEI_IP} route-map {RM_NAME} in
\end{codebox}

This instance makes \texttt{agg-p0-a1} the preferred next hop, shifting the flow from \texttt{agg-p0-a0}/core group~0 to \texttt{agg-p0-a1}/core group~1.

\begin{codebox}[LP instance: on edge-p0-e0, shift the flow to agg-p0-a1]
route-map RM_LP_AGG1 permit 10
 set local-preference 300
route-map RM_LP_AGG1 permit 20
!
router bgp 100000
 address-family ipv4
  neighbor 2.0.1.1 route-map RM_LP_AGG1 in
\end{codebox}

\noindent\textbf{Multi-Exit Discriminator (MED).}
\noindent The template is filled with: (i) \texttt{ASN}: the router's own ASN; (ii) \texttt{NEI\_IP}: the link IP of the neighbor whose entry path is being penalized; (iii) \texttt{MED\_HIGH}: any large value (lower MED is preferred in BGP). \textbf{\emph{The effect:}} routes received from \texttt{NEI\_IP} carry a high MED, so among all the parallel links between the two ASes, the neighbor will pick a different (lower-MED) link to send the flow over.

\begin{codebox}[MED template]
route-map {RM_NAME} permit 10
 set metric {MED_HIGH}
route-map {RM_NAME} permit 20
!
router bgp {ASN}
 address-family ipv4
  neighbor {NEI_IP} route-map {RM_NAME} in
\end{codebox}

This instance penalizes routes from \texttt{core-g0-c0}, so \texttt{agg-p0-a0} shifts the flow to \texttt{core-g0-c1} within the same core group.

\begin{codebox}[MED instance: on agg-p0-a0, push traffic away from core-g0-c0]
route-map RM_MED_CORE0 permit 10
 set metric 9999
route-map RM_MED_CORE0 permit 20
!
router bgp 20000
 address-family ipv4
  neighbor 3.0.0.1 route-map RM_MED_CORE0 in
\end{codebox}

\noindent\textbf{Static route.}
\noindent The template is filled with: (i) \texttt{DST\_NET}/\texttt{DST\_MASK}: the destination prefix of the queried flow; (ii) \texttt{NEXT\_HOP\_IP}: the link IP of the chosen neighbor (Table~\ref{tab:ip-scheme}). \textbf{\emph{The effect:}} any packet matching \texttt{DST\_NET} is forwarded to \texttt{NEXT\_HOP\_IP}, overriding the BGP-selected next hop. 
\begin{codebox}[Static route template]
ip route {DST_NET} {DST_MASK} {NEXT_HOP_IP}
\end{codebox}

This instance forces \texttt{agg-p0-a0} to send traffic for \texttt{1.5.3.0/24} directly to \texttt{core-g0-c5}, overriding its BGP choice.

\begin{codebox}[Static route instance: on agg-p0-a0, pin the destination to core-g0-c5]
ip route 1.5.3.0 255.255.255.0 3.0.5.1
\end{codebox}

\noindent\textbf{BGP network advertisement.}
\noindent The template is filled with: (i) \texttt{ASN}: the router's own ASN; (ii) \texttt{DST\_NET}/\texttt{DST\_MASK}: a prefix more specific than the default fabric aggregate; (iii) \texttt{NEXT\_HOP\_IP}: a neighbor IP that anchors this prefix in the local routing table. The \texttt{ip route} line provides the backing route required by Cisco~IOS before a \texttt{network} statement can originate the prefix. \textbf{\emph{The effect:}} this router announces \texttt{DST\_NET} into BGP, and because longest-prefix match wins, peers steer the corresponding flow to it.

\begin{codebox}[BGP network template]
ip route {DST_NET} {DST_MASK} {NEXT_HOP_IP}
!
router bgp {ASN}
 address-family ipv4
  network {DST_NET} mask {DST_MASK}
\end{codebox}

This instance makes \texttt{agg-p0-a1} advertise the more-specific \texttt{1.5.3.0/24}, attracting the flow to \texttt{agg-p0-a1} via longest-prefix match.

\begin{codebox}[BGP network instance: on agg-p0-a1, originate 1.5.3.0/24]
ip route 1.5.3.0 255.255.255.0 3.0.0.3
!
router bgp 20000
 address-family ipv4
  network 1.5.3.0 mask 255.255.255.0
\end{codebox}

\noindent\textbf{Prefix aggregation.}
\noindent The fields are filled in the same way as the BGP-network template above; the only addition is the \texttt{aggregate-address} line on the same prefix. \textbf{\emph{The effect:}} any flow falling inside \texttt{DST\_NET} loses its individual paths and is forwarded along the aggregated route originated by this router.

\begin{codebox}[Prefix aggregation template]
ip route {DST_NET} {DST_MASK} {NEXT_HOP_IP}
!
router bgp {ASN}
 address-family ipv4
  network {DST_NET} mask {DST_MASK}
  aggregate-address {DST_NET} {DST_MASK} as-set
\end{codebox}

This instance makes \texttt{agg-p0-a1} originate and aggregate \texttt{1.5.3.0/24}, causing matching traffic to follow \texttt{agg-p0-a1}'s aggregated route.

\begin{codebox}[Prefix aggregation instance: on agg-p0-a1, aggregate 1.5.3.0/24]
ip route 1.5.3.0 255.255.255.0 3.0.0.3
!
router bgp 20000
 address-family ipv4
  network 1.5.3.0 mask 255.255.255.0
  aggregate-address 1.5.3.0 255.255.255.0 as-set
\end{codebox}

\noindent\textbf{Interface shutdown.}
\noindent The template is filled with one field, \texttt{INTF\_NAME}, the name of an interface on a router along the queried path. \textbf{\emph{The effect:}} the corresponding link disappears from the topology, so any flow that previously traversed it must take an alternate route.

\begin{codebox}[Interface shutdown template]
interface {INTF_NAME}
 shutdown
\end{codebox}

This instance disables the default link from \texttt{edge-p0-e0} to \texttt{agg-p0-a0}, forcing affected traffic onto another available aggregation uplink.

\begin{codebox}[Interface shutdown instance: on edge-p0-e0, disable the link toward agg-p0-a0]
interface GigabitEthernet1/0/3
 shutdown
\end{codebox}

\paragraph{Change instantiation.}
For each flow, we first identify the configuration template that currently controls its forwarding behavior, then inject the change by appending the new template instance before the device configuration's trailing \texttt{end}. This append-only procedure mirrors how operators apply incremental updates through command-line configuration tools: the newly added commands override the earlier template for the affected flow without rewriting the original configuration file.

\section{Agent Design Details}
\label{app:workflow}

Each agent consists of a \textbf{guidance prompt} and a \textbf{execution script}; \methodagent{} additionally uses skill documents and tools.
The prompt combines shared task instructions (reasoning, schema, examples, and references) with an agent-specific \texttt{AGENT.md} that defines its evidence-access workflow.
The script orchestrates LLM calls, parses JSON actions, dispatches operations (configuration retrieval, tool calls, or skill loading), and enforces interaction limits; agents share this structure but differ in action vocabulary and dispatch logic.
All agents are evaluated by a common framework that loads ground-truth instances, runs each agent, and compares normalized path sequences.

\subsection{Shared Guidance Prompt}
\label{app:shared}

The shared guidance prompt provides four categories of information that ground agentic path-analysis tasks regardless of which agent architecture is used. We summarize them below.

\begin{itemize}[leftmargin=*,itemsep=4pt,topsep=4pt]
\item \textbf{Input/output schema.}
The prompt defines the task (analyze before/after forwarding paths for a source--destination pair given a configuration change) and specifies the input and output format. The input provides source and destination edge routers, the queried prefixes, and a list of configuration diffs. 
If source or destinaion prefixes are not specified, we default them to ``any'' (\texttt{0.0.0.0/0}).
The output is a JSON object:

\begin{codebox}[Input Schema (JSON)]
{"query": {"src": "edge-p0-e0",
           "dst": "edge-p1-e0",
           "prefix": "1.1.0.0/24"},
 "change": [{"device": "edge-p0-e0",
   "add_lines": ["route-map RM_LP_AGG1 permit 10",
     " set local-preference 300",
     "router bgp 100000",
     " address-family ipv4",
     "  neighbor 2.0.1.0 route-map RM_LP_AGG1 in"
   ],
   "description": "LP=300 on edge inbound from agg1"}]}
\end{codebox}

\begin{codebox}[Output Schema (JSON)]
{
  "before_path": ["edge-p0-e0", "agg-p0-a0",
    "core-g0-c0", "agg-p1-a0", "edge-p1-e0"],
  "after_path":  ["edge-p0-e0", "agg-p0-a1",
    "core-g1-c0", "agg-p1-a1", "edge-p1-e0"],
  "decisive_evidence": "LP=300 from agg1 overrides
    LP=200 from agg0 on edge-p0-e0. Core group shifts from 0 to 1."
}
\end{codebox}

\item \textbf{Chain-of-thought (CoT) guidance.}
All agents embed a structured hop-by-hop reasoning skeleton that instructs the LLM to trace the forwarding path step by step:

\begin{codebox}[CoT Reasoning Skeleton]
1. Identify source and destination edge routers.
2. BEFORE state: trace hop-by-hop from src_edge.
   At each router:
   - Check static routes (AD=1 overrides BGP AD=20)
  - Apply longest-prefix match (/24 > /16 > /8)
   - Check BGP: route-map -> LP from each neighbor
   - If no LP set, check MED; if no MED, lowest
     router-ID tiebreak
   - agg index = core group = dst_agg index
3. AFTER state: apply the config change, re-trace
   with the same logic.
4. Report both paths and the decisive evidence.
\end{codebox}

\noindent The skeleton encodes the route-selection priority (static AD$=$1 beats BGP, longest-prefix match precedes BGP tie-breaking; within BGP: highest LP $>$ shortest AS-path $>$ lowest MED $>$ lowest router-ID) and the fat-tree structural invariant that the aggregation index determines the core group and destination aggregation.

\item \textbf{Few-shot demonstrations.}
Each prompt includes worked examples on a small topology covering common analysis patterns and preventing recurring errors. For intra-datacenter fat-tree, five examples demonstrate LP on edge (shifting agg), LP on agg (shifting core), interface shutdown, BGP network with longest-prefix match, and static route overriding BGP. A representative example:

\begin{codebox}[Few-shot Example: LP on edge shifts agg]
Query: src=edge-p0-e0, dst=edge-p1-e0, prefix=1.1.0.0/24
Change: On edge-p0-e0, add route-map RM_LP_AGG1
  with set local-preference 300, applied inbound
  from agg-p0-a1.

Analysis:
- BEFORE: edge has LP=200 from agg0 (RM_PIN),
  LP=100(default) from agg1. Picks agg0 (200>100).
  agg0 -> core-g0-c0 (lowest RID). dst: agg-p1-a0.
- AFTER: LP=300 from agg1, LP=200 from agg0.
  Picks agg1 (300>200). agg1 -> core-g1-c0. dst:
  agg-p1-a1.
Answer:
  before: [edge-p0-e0, agg-p0-a0, core-g0-c0,
           agg-p1-a0, edge-p1-e0]
  after:  [edge-p0-e0, agg-p0-a1, core-g1-c0,
           agg-p1-a1, edge-p1-e0]
\end{codebox}

\noindent These examples are chosen to demonstrate non-obvious patterns (e.g., /24 beating /8 via longest-prefix match regardless of LP; static route AD$=$1 overriding all BGP attributes) that LLMs tend to get wrong without explicit guidance.

\item \textbf{Reference material.}
Following common agentic-framework practice, the prompt provides background reference on the specific topology structure (device naming, connectivity rules, path length) and protocol details (the six configuration mechanisms and their routing effects).
\end{itemize}

\paragraph{Task-specific variations.}
The shared prompt is tailored for each of the three task types---intra-datacenter fat-tree, inter-datacenter, and compound changes---with the differences concentrated in the reference material and few-shot examples.
Inter-datacenter prompts describe the dual-datacenter topology (device prefixes \texttt{dc0-}/\texttt{dc1-}, gateway interconnection) and include examples demonstrating that changes in datacenter-0 do not affect datacenter-1's routing.
Compound-change prompts use the same intra-datacenter topology reference but include examples with 2--4 simultaneous changes, illustrating how to analyze each change independently and then compose their combined effect on the path.

\subsection{\methodnaive{}}
\label{app:full-context}

The \methodnaive{} agent places the shared guidance, full topology, all router configurations, query, and configuration change into one prompt.
Its minimal \texttt{AGENT.md} asks the model to return the JSON answer in a single LLM call, with no follow-up interaction.

\subsection{\methodmulti{}}
\label{app:multi-round}

The \methodmulti{} agent starts with the full topology and query but no device configurations.
Across up to $K{=}10$ rounds, it requests device files via \texttt{request\_configs}; the script returns the relevant \texttt{.cfg} files and iterates until the model answers or the round limit is reached.

\subsection{\methodagent{}}
\label{app:agent-skills}

\subsubsection{Agent Workflow}

\paragraph{Agent-specific guidance.}
The \methodagent{} \texttt{AGENT.md} defines a plan-invoke-reason loop and lists the 14~available skills by category (configuration extraction, topology, and verification).
The model selects a skill with \texttt{\{"action": "use\_skill", "skill": "skill-name"\}}, receives its \texttt{SKILL.md} workflow, and then calls tools as JSON arrays \texttt{[\{"tool": "tool\_name", "args": \{...\}\}, ...]}.
Before an answer is accepted, the runtime enforces a device-inspection requirement that grows with network size.
The \texttt{AGENT.md} also includes accumulated don't rules (\Cref{app:dont-rules}) to prevent recurring topology and tool-use errors.
The initial prompt provides no topology graph or device configurations; all evidence must be acquired through skill and tool invocations.

\paragraph{Execution script.}
The execution script deterministically parses each JSON action: \texttt{use\_skill} returns the requested \texttt{SKILL.md}, tool-call arrays execute stateless scripts over the network files, and \texttt{answer} terminates the loop.
The loop stops on a valid answer or after $S{=}100$ steps, then forces a final answer with up to 3 retries for output format errors.

\subsubsection{Skills and Tools}

Skills and tools form a two-layer architecture that separates domain reasoning from mechanical data access.

\paragraph{Tools.}
A \textbf{tool} is a stateless, deterministic script that performs a narrow operation on network data (topology JSON or device \texttt{.cfg} files) and returns structured, focused results.
Nine tools are available (Table~\ref{tab:tools}).
Tools perform no reasoning---they search, parse, and extract only the evidence requested.

\begin{table}[t]
\centering
\caption{Low-level tools available to skill workflows.}
\label{tab:tools}
\small
\renewcommand{\arraystretch}{1.15}
\resizebox{\textwidth}{!}{
\begin{tabular}{@{}p{6.2cm}p{6.8cm}@{}}
\toprule
Tool & Description \\
\midrule
\toolname{search\_config(device, regex)} & Regex search across a device's \texttt{.cfg} file. Returns only matching lines with surrounding context. \\
\toolname{read\_config\_block(device, block\_type, name)} & Extract a named configuration block (e.g., \texttt{route-map}, \texttt{prefix-list}, \texttt{router bgp}, \texttt{community-list}). \\
\toolname{get\_bgp\_neighbors(device)} & Parse the \texttt{router bgp} section and return configured BGP neighbors with applied route-maps (in/out). \\
\toolname{get\_interface\_config(device, intf)} & Return interface IP address, shutdown status, and description. \\
\toolname{get\_topology\_adjacency(device)} & From the topology JSON, return all neighbors: \texttt{(neighbor, local\_intf, remote\_intf)}. \\
\toolname{get\_layer\_devices(layer, scope)} & Return devices in a synthetic fat-tree layer (edge/agg/core), optionally scoped to a pod. \\
\toolname{get\_topology\_paths(src, dst)} & Enumerate topology-valid candidate paths between two endpoints. \\
\toolname{find\_references(device, symbol)} & Return configuration lines that reference a named object (route-map, prefix-list, community-list, interface). \\
\toolname{extract\_match\_set\_clauses(device, route\_map)} & Return the \texttt{match} and \texttt{set} clauses of a route-map, grouped by sequence number. \\
\bottomrule
\end{tabular}}
\end{table}

\paragraph{Skills.}
A \textbf{skill} is a reusable multi-step workflow specification stored as a \texttt{SKILL.md} document.
Each skill defines a high-level objective, describes the reasoning chain from objective to evidence, lists which tools to call at each step, and specifies the structured result to return.
The 14~skills are organized into four categories.
The critical value of skills is that they encode the \emph{domain reasoning chain} connecting a high-level question to scattered configuration evidence---for instance, from ``LP for prefix~$P$'' to BGP neighbors, inbound route-maps, and their match/set clauses.

\paragraph{Config extraction skills (7 skills).}
These skills navigate the indirect, reference-heavy structure of device configurations to extract specific routing attributes.

\begin{itemize}[leftmargin=*,itemsep=4pt,topsep=4pt]
\item \toolname{find-lp-for-prefix(device, prefix)}:
LP is typically applied indirectly via a route-map attached to a BGP neighbor.
The workflow is: (1)~call \toolname{get\_bgp\_neighbors} to find peers and inbound route-maps; (2)~call \toolname{extract\_match\_set\_clauses} on each route-map; (3)~resolve referenced prefix-lists if needed; (4)~report any \texttt{set local-preference} values, or default LP$=$100.

\item \toolname{find-med-for-prefix(device, prefix)}:
Follows the same route-map workflow as LP, but extracts \texttt{set metric} clauses and reports MED values.

\item \toolname{find-static-route(device, prefix)}:
Call \toolname{search\_config} for \texttt{ip route} entries matching the prefix and parse the network, mask, and next-hop IP.

\item \toolname{find-bgp-network(device, prefix)}:
Call \toolname{search\_config} for \texttt{network} entries matching the prefix.

\item \toolname{find-prefix-aggregation(device, prefix)}:
Search for \texttt{aggregate-address} entries and report the configured aggregates and whether \texttt{summary-only} is present.

\item \toolname{find-route-map-policy(device, neighbor\_ip)}:
Call \toolname{get\_bgp\_neighbors} to locate the target peer by IP, then use \toolname{extract\_match\_set\_clauses} to return its inbound and outbound route-map clauses.

\item \toolname{check-interface-reachability(device, intf)}:
Call \toolname{get\_interface\_config(device, intf)} to check IP assignment and shutdown status.
\end{itemize}

\paragraph{Topology extraction skills (3 skills).}
These skills expose network structure at different levels of abstraction.

\begin{itemize}[leftmargin=*,itemsep=4pt,topsep=4pt]
\item \toolname{find-neighbors(device)}:
Call \toolname{get\_topology\_adjacency(device)} to list all neighbors; optionally call \toolname{get\_interface\_config} on each to enrich with IP and status information.

\item \toolname{find-layer-devices(layer, scope)}:
Call \toolname{get\_layer\_devices(layer, scope)} to return devices in the specified synthetic fat-tree layer and optional pod scope.

\item \toolname{find-candidate-paths(src, dst, prefix)}:
Call \toolname{get\_topology\_paths(src, dst)} and keep paths with valid hop counts (3 for intra-pod, 5 for inter-pod, 8 for cross-datacenter).
\end{itemize}

\paragraph{Config verification skills (2 skills).}
These skills answer higher-level decision questions by composing extraction skills and applying routing protocol knowledge.

\begin{itemize}[leftmargin=*,itemsep=4pt,topsep=4pt]
\item \toolname{determine-route-selection(device, prefix)}:
Compose static-route, BGP-network, LP, MED, aggregation, and neighbor-status evidence, then report the winning route and next-hop device.

\item \toolname{verify-bgp-session(device, neighbor)}:
Call \toolname{get\_bgp\_neighbors} on both devices and report whether both sides expose BGP neighbor state.
\end{itemize}

\paragraph{Topology verification skills (2 skills).}
These skills verify physical-layer reachability.

\begin{itemize}[leftmargin=*,itemsep=4pt,topsep=4pt]
\item \toolname{verify-link(device1, device2)}:
Call \toolname{get\_topology\_adjacency} to confirm the link exists; call \toolname{get\_interface\_config} on both ends to check that neither interface is administratively shut down.

\item \toolname{verify-path-reachability(path)}:
For each consecutive pair of routers in \texttt{path}, invoke \toolname{verify-link}; report the first broken link if any.
\end{itemize}

\subsubsection{Don't Rules}
\label{app:dont-rules}

As our primary agent design, \methodagent{} differs fundamentally from the other two agents in that it starts with \emph{no initial network information}---no topology graph, no device configurations.
All evidence must be acquired through skill and tool invocations, which means the agent's accuracy depends entirely on its planning quality: which devices to query, which routing attributes to check, and how to interpret the results.
During development, we observed that the agent's errors are largely caused by two categories of uncontrolled variables: \emph{topology-induced} failures (misunderstanding the fat-tree connectivity structure) and \emph{tool-induced} failures (misusing tool APIs or misinterpreting tool results).
These failures are not inherent to the agent's reasoning capability; rather, they stem from unfamiliarity with benchmark-specific topology conventions already provided to the other agentic methods and with tool interfaces shaped by the initial prompt design.

To isolate the agent's true analytical capability from such confounding factors, we added a set of don't rules that encode recurring anti-patterns observed during development:
\begin{itemize}[leftmargin=*,itemsep=3pt,topsep=3pt]
\item \textbf{Topology-induced rules} specify fat-tree connectivity facts that the agent often infers incorrectly. For example, the selected core is determined by the lowest router-ID within a core group, not by matching the core index to the aggregation index; inter-pod paths must also satisfy \texttt{src\_agg index == core group == dst\_agg index}.
\item \textbf{Tool-induced rules} specify how to use and interpret tool outputs. For example, \texttt{find-candidate-paths} only enumerates topology-valid paths and does not account for interface shutdowns, so shutdown changes require explicit reachability checks; other rules prevent incorrect tool argument names.
\end{itemize}
We also include output-consistency rules, such as re-reading \texttt{decisive\_evidence} against the JSON paths before submitting and avoiding router-ID tiebreaks when a static route deterministically overrides BGP.
These rules eliminate confounding variables and ensure that evaluation measures the agent's path-analysis capability rather than its familiarity with benchmark-specific conventions.

\section{Complete Experimental Results}
\label{app:all-results}

This appendix gives the per-LLM, per-$k$, per-agentic-method results introduced in~\Cref{sec:results}, plus the concrete performance-metric values of the real-world hyperscale evaluation. Overall, these complete results support all claims made in \Cref{sec:results}.

\subsection{Real-World Hyperscale Evaluation}
\label{app:results-realnet}

Table~\ref{tab:realnet-gpt54} reports the performance values from our real-world hyperscale evaluation, complementing \Cref{sec:realnet}. The accuracy closely aligns with \benchmarkname{}, showing that \benchmarkname{} can effectively guide agent design for real production networks.

\begin{table}[H]
\centering
\small
\caption{\methodagent{} (GPT-5.4) performance on the real-world production network reported in \Cref{sec:realnet}.}
\label{tab:realnet-gpt54}
\resizebox{0.7\linewidth}{!}{%
\begin{tabular}{@{}llrrrrr@{}}
\toprule
Scenario & Devices & Accuracy & Tokens & $t_{\mathrm{LLM}}$ & $t_{\mathrm{Total}}$ & $R_{\mathrm{LLM}}$ \\
\midrule
Intra-datacenter & ${\sim}$10K &\accM{92\%}&\tokM{236K}&\tlH{ 719 s}&\rtM{1023 s}&\rndVL{5.9}\\
Inter-datacenter & ${\sim}$20K &\accLM{73\%}&\tokH{424K}&\tlH{1293 s}&\rtH{1762 s}&\rndL{7.5}\\
\bottomrule
\end{tabular}
}
\end{table}

\subsection{Three-Agent and Six-LLM Comparison on Intra-Datacenter Single-Change Analysis}
\label{app:results-ft}

We present the complete results of the three agents under all six LLMs on intra-datacenter single-change analysis across all scales (from $k{=}4$ to $k{=}200$) in Tables~\ref{tab:app-ft-gpt-5-4}--\ref{tab:app-ft-llama}. These results further validate the design principle from \Cref{sec:overall}: \emph{``explore more; digest less.''}

For \methodnaive{} and \methodmulti{}, all six LLMs still suffer severe cost-side scalability issues: tokens and runtime grow rapidly, and both methods hit context-window overflow or the time limit at small scales. Specifically, \methodnaive{} stops scaling at $k{=}16$ (320 routers) on the strongest LLMs (GPT-5.4 and DeepSeek-V4-Pro) and at $k{=}8$ (80 routers) on all the others (GPT-4o, OpenAI o3, Qwen-3.5-397B, and Llama-3.3-70B); \methodmulti{} stops at $k{=}20$ (500 routers) on most LLMs and at $k{=}16$ (320 routers) on GPT-4o and Llama-3.3-70B. On the accuracy side, both methods also degrade severely with scale on weaker LLMs (e.g., GPT-4o and Llama-3.3-70B) and on those that tend to overthink (e.g., Qwen-3.5-397B).

\methodagent{}, in contrast, scales to at least $k{=}120$ (18K routers) on every LLM, demonstrating its strong scalability. Except for the weakest LLM (Llama-3.3-70B), all LLMs keep their accuracy stable (mostly ${>}90\%$); GPT-5.4 and the thinking models (OpenAI o3, DeepSeek-V4-Pro, and Qwen-3.5-397B) approach $100\%$, and token consumption stays nearly flat across scales---far less explosive than under the other two methods. The thinking models, however, exhibit instability relative to the general-purpose GPT-5.4: they sometimes spiral into excessive internal reasoning on individual cases, causing token/runtime spikes (e.g., OpenAI o3 shows an abnormal token spike at $k{=}16$, jumping to $272$K from $123$K at $k{=}12$ and $134$K at $k{=}20$, which we traced to unusually long thinking on a few random cases). Llama-3.3-70B, the weakest model, fluctuates in both accuracy and tokens at every scale, with some correct answers coming from guessing rather than reasoning. Across all scales, GPT-5.4 is the most stable in both accuracy and token cost, confirming the agent design principle from \Cref{sec:overall}: \emph{``explore more; digest less.''}

\begin{table}[H]
\centering
\caption{Three-agent comparison on intra-datacenter fat-tree with GPT-5.4 across all scales.}
\label{tab:app-ft-gpt-5-4}
\footnotesize
\setlength{\tabcolsep}{4pt}
\begin{tabular}{@{}llrrrr!{\hspace{6pt}}rrrr!{\hspace{6pt}}rrrr@{}}
\toprule
$k$ & Routers & \multicolumn{4}{c}{\textbf{\methodnaive{}}} & \multicolumn{4}{c}{\textbf{\methodmulti{}}} & \multicolumn{4}{c}{\textbf{\methodagent{}}} \\
\cmidrule(lr){3-6}\cmidrule(lr){7-10}\cmidrule(lr){11-14}
~ & ~ & Acc & Tok & $t_\mathrm{Run}$ & $R_\mathrm{LLM}$ & Acc & Tok & $t_\mathrm{Run}$ & $R_\mathrm{LLM}$ & Acc & Tok & $t_\mathrm{Run}$ & $R_\mathrm{LLM}$ \\
\midrule
4 & 20    &\accH{99\%}&\tokVL{20K}&\rtVL{66s}&\rndVL{1.0}&\accH{100\%}&\tokVL{22K}&\rtVL{87s}&\rndVL{3.2}&\accH{100\%}&\tokVL{48K}&\rtVL{187s}&\rndVL{6.9}\\
8 & 80    &\accH{98\%}&\tokL{100K}&\rtVL{317s}&\rndVL{1.0}&\accH{100\%}&\tokVL{50K}&\rtVL{189s}&\rndVL{3.1}&\accH{98\%}&\tokVL{51K}&\rtVL{236s}&\rndVL{6.8}\\
12 & 180  &\accM{94\%}&\tokM{292K}&\rtL{905s}&\rndVL{1.0}&\accH{99\%}&\tokL{114K}&\rtVL{388s}&\rndVL{3.0}&\accH{100\%}&\tokVL{51K}&\rtVL{246s}&\rndVL{6.7}\\
16 & 320  &\accH{99\%}&\tokH{644K}&\rtH{1982s}&\rndVL{1.0}&\accHM{96\%}&\tokM{234K}&\rtL{752s}&\rndVL{3.0}&\accH{100\%}&\tokVL{52K}&\rtVL{243s}&\rndVL{6.7}\\
20 & 500  & \multicolumn{4}{c}{Context Window Overflow}                                          &\accH{100\%}&\tokH{427K}&\rtM{1473s}&\rndVL{2.9}&\accH{100\%}&\tokVL{53K}&\rtL{554s}&\rndVL{6.8}\\
30 & 1{,}125 & & & & & \multicolumn{4}{c}{Time Limit Exceeded}                                  &\accH{100\%}&\tokVL{55K}&\rtL{554s}&\rndVL{6.8}\\
50 & 3{,}125 & & & & & & & & &\accH{99\%}&\tokVL{67K}&\rtL{663s}&\rndL{7.7}\\
80 & 8{,}000 & & & & & & & & &\accH{99\%}&\tokVL{68K}&\rtL{592s}&\rndL{7.5}\\
120 & 18{,}000 & & & & & & & & &\accH{99\%}&\tokVL{89K}&\rtL{903s}&\rndL{8.8}\\
160 & 32{,}000 & & & & & & & & &\accH{100\%}&\tokL{112K}&\rtH{1630s}&\rndM{10.2}\\
200 & 50{,}000 & & & & & & & & &\accH{99\%}&\tokL{119K}&\rtH{2754s}&\rndM{10.4}\\
\bottomrule
\end{tabular}
\end{table}

\begin{table}[H]
\centering
\caption{Three-agent comparison on intra-datacenter fat-tree with GPT-4o across all scales.}
\label{tab:app-ft-gpt-4o}
\scriptsize
\setlength{\tabcolsep}{2.5pt}
\resizebox{\textwidth}{!}{%
\begin{tabular}{@{}llrrrr!{\hspace{10pt}}rrrr!{\hspace{10pt}}rrrr@{}}
\toprule
$k$ & Routers & \multicolumn{4}{c}{\textbf{\methodnaive{}}} & \multicolumn{4}{c}{\textbf{\methodmulti{}}} & \multicolumn{4}{c}{\textbf{\methodagent{}}} \\
~ & ~ & Acc & Tok & $t_\mathrm{Run}$ & $R_\mathrm{LLM}$ & Acc & Tok & $t_\mathrm{Run}$ & $R_\mathrm{LLM}$ & Acc & Tok & $t_\mathrm{Run}$ & $R_\mathrm{LLM}$ \\
\midrule
4 & 20 &\accLM{78\%}&\tokVL{19K}&\rtVL{26s}&\rndVL{1.0}&\accM{94\%}&\tokVL{26K}&\rtVL{46s}&\rndVL{3.4}&\accM{93\%}&\tokL{98K}&\rtVL{175s}&\rndM{12.9}\\
8 & 80 &\accL{67\%}&\tokL{100K}&\rtVL{113s}&\rndVL{1.0}&\accL{67\%}&\tokVL{56K}&\rtVL{98s}&\rndVL{3.5}&\accM{91\%}&\tokL{101K}&\rtVL{250s}&\rndM{12.7}\\
12 & 180 & \multicolumn{4}{c}{Context Window Overflow} &\accL{66\%}&\tokM{140K}&\rtVL{191s}&\rndVL{3.7}&\accM{89\%}&\tokL{105K}&\rtVL{278s}&\rndM{12.9}\\
16 & 320 &  &  &  &  &\accL{52\%}&\tokM{283K}&\rtVL{334s}&\rndVL{3.6}&\accM{89\%}&\tokL{117K}&\rtVL{294s}&\rndM{13.7}\\
20 & 500 &  &  &  &  & \multicolumn{4}{c}{Context Window Overflow} &\accH{98\%}&\tokL{117K}&\rtL{894s}&\rndM{13.4}\\
30 & 1,125 &  &  &  &  &  &  &  &  &\accHM{96\%}&\tokM{168K}&\rtM{1029s}&\rndM{15.1}\\
50 & 3,125 &  &  &  &  &  &  &  &  &\accHM{96\%}&\tokM{235K}&\rtM{1337s}&\rndM{18.6}\\
80 & 8,000 &  &  &  &  &  &  &  &  &\accHM{96\%}&\tokM{230K}&\rtM{1187s}&\rndM{18.6}\\
120 & 18,000 &  &  &  &  &  &  &  &  &\accM{95\%}&\tokM{351K}&\rtH{2026s}&\rndH{23.3}\\
160 & 32,000 &  &  &  &  &  &  &  &  & \multicolumn{4}{c}{Time Limit Exceeded} \\
200 & 50,000 &  &  &  &  &  &  &  &  &  &  &  &  \\
\bottomrule
\end{tabular}}
\end{table}

\begin{table}[H]
\centering
\caption{Three-agent comparison on intra-datacenter fat-tree with OpenAI o3 across all scales.}
\label{tab:app-ft-o3}
\scriptsize
\setlength{\tabcolsep}{2.5pt}
\resizebox{\textwidth}{!}{%
\begin{tabular}{@{}llrrrr!{\hspace{10pt}}rrrr!{\hspace{10pt}}rrrr@{}}
\toprule
$k$ & Routers & \multicolumn{4}{c}{\textbf{\methodnaive{}}} & \multicolumn{4}{c}{\textbf{\methodmulti{}}} & \multicolumn{4}{c}{\textbf{\methodagent{}}} \\
~ & ~ & Acc & Tok & $t_\mathrm{Run}$ & $R_\mathrm{LLM}$ & Acc & Tok & $t_\mathrm{Run}$ & $R_\mathrm{LLM}$ & Acc & Tok & $t_\mathrm{Run}$ & $R_\mathrm{LLM}$ \\
\midrule
4 & 20 &\accH{98\%}&\tokVL{22K}&\rtVL{40s}&\rndVL{1.0}&\accH{99\%}&\tokVL{23K}&\rtVL{52s}&\rndVL{2.8}&\accH{99\%}&\tokL{112K}&\rtVL{247s}&\rndM{12.7}\\
8 & 80 &\accM{94\%}&\tokL{102K}&\rtVL{170s}&\rndVL{1.0}&\accH{99\%}&\tokVL{42K}&\rtVL{94s}&\rndVL{2.5}&\accH{97\%}&\tokL{98K}&\rtVL{291s}&\rndM{12.0}\\
12 & 180 & \multicolumn{4}{c}{Context Window Overflow} &\accH{98\%}&\tokVL{85K}&\rtVL{162s}&\rndVL{2.3}&\accM{93\%}&\tokL{123K}&\rtVL{361s}&\rndM{12.8}\\
16 & 320 &  &  &  &  &\accH{98\%}&\tokM{156K}&\rtVL{268s}&\rndVL{2.0}&\accHM{96\%}&\tokM{272K}&\rtL{697s}&\rndH{21.5}\\
20 & 500 &  &  &  &  &\accHM{96\%}&\tokM{272K}&\rtL{530s}&\rndVL{1.9}&\accH{98\%}&\tokL{134K}&\rtL{984s}&\rndM{13.4}\\
30 & 1,125 &  &  &  &  & \multicolumn{4}{c}{Context Window Overflow} &\accH{97\%}&\tokM{238K}&\rtM{1495s}&\rndM{19.8}\\
50 & 3,125 &  &  &  &  &  &  &  &  &\accM{95\%}&\tokM{318K}&\rtH{1971s}&\rndH{25.0}\\
80 & 8,000 &  &  &  &  &  &  &  &  &\accH{97\%}&\tokM{282K}&\rtH{1585s}&\rndH{22.4}\\
120 & 18,000 &  &  &  &  &  &  &  &  &\accH{98\%}&\tokM{244K}&\rtH{1758s}&\rndM{19.4}\\
160 & 32,000 &  &  &  &  &  &  &  &  &\accH{98\%}&\tokM{262K}&\rtH{3078s}&\rndH{21.1}\\
200 & 50,000 &  &  &  &  &  &  &  &  & \multicolumn{4}{c}{Time Limit Exceeded} \\
\bottomrule
\end{tabular}}
\end{table}

\begin{table}[H]
\centering
\caption{Three-agent comparison on intra-datacenter fat-tree with DeepSeek-V4-Pro across all scales.}
\label{tab:app-ft-deepseek-v4-pro}
\scriptsize
\setlength{\tabcolsep}{2.5pt}
\resizebox{\textwidth}{!}{%
\begin{tabular}{@{}llrrrr!{\hspace{10pt}}rrrr!{\hspace{10pt}}rrrr@{}}
\toprule
$k$ & Routers & \multicolumn{4}{c}{\textbf{\methodnaive{}}} & \multicolumn{4}{c}{\textbf{\methodmulti{}}} & \multicolumn{4}{c}{\textbf{\methodagent{}}} \\
~ & ~ & Acc & Tok & $t_\mathrm{Run}$ & $R_\mathrm{LLM}$ & Acc & Tok & $t_\mathrm{Run}$ & $R_\mathrm{LLM}$ & Acc & Tok & $t_\mathrm{Run}$ & $R_\mathrm{LLM}$ \\
\midrule
4 & 20 &\accHM{96\%}&\tokVL{24K}&\rtVL{80s}&\rndVL{1.0}&\accHM{96\%}&\tokVL{34K}&\rtVL{125s}&\rndVL{3.3}&\accHM{96\%}&\tokM{144K}&\rtL{520s}&\rndM{13.5}\\
8 & 80 &\accM{87\%}&\tokL{108K}&\rtVL{343s}&\rndVL{1.0}&\accH{97\%}&\tokVL{67K}&\rtVL{245s}&\rndVL{3.3}&\accH{98\%}&\tokM{160K}&\rtL{662s}&\rndM{15.0}\\
12 & 180 &\accM{90\%}&\tokM{309K}&\rtL{957s}&\rndVL{1.0}&\accH{99\%}&\tokM{141K}&\rtL{472s}&\rndVL{3.2}&\accH{99\%}&\tokM{161K}&\rtL{685s}&\rndM{14.8}\\
16 & 320 &\accLM{82\%}&\tokH{674K}&\rtH{2071s}&\rndVL{1.0}&\accH{97\%}&\tokM{260K}&\rtL{831s}&\rndVL{3.0}&\accH{100\%}&\tokM{149K}&\rtL{641s}&\rndM{14.4}\\
20 & 500 & \multicolumn{4}{c}{Context Window Overflow} &\accH{99\%}&\tokH{497K}&\rtH{1701s}&\rndVL{3.1}&\accH{100\%}&\tokM{171K}&\rtM{1445s}&\rndM{15.9}\\
30 & 1,125 &  &  &  &  & \multicolumn{4}{c}{Time Limit Exceeded} &\accH{100\%}&\tokM{162K}&\rtM{1355s}&\rndM{15.1}\\
50 & 3,125 &  &  &  &  &  &  &  &  &\accH{99\%}&\tokM{184K}&\rtH{1512s}&\rndM{16.0}\\
80 & 8,000 &  &  &  &  &  &  &  &  &\accH{100\%}&\tokM{195K}&\rtM{1458s}&\rndM{16.8}\\
120 & 18,000 &  &  &  &  &  &  &  &  &\accH{99\%}&\tokM{214K}&\rtH{1912s}&\rndM{17.6}\\
160 & 32,000 &  &  &  &  &  &  &  &  &\accH{99\%}&\tokM{237K}&\rtH{3124s}&\rndM{19.0}\\
200 & 50,000 &  &  &  &  &  &  &  &  & \multicolumn{4}{c}{Time Limit Exceeded} \\
\bottomrule
\end{tabular}}
\end{table}

\begin{table}[H]
\centering
\caption{Three-agent comparison on intra-datacenter fat-tree with Qwen-3.5-397B across all scales.}
\label{tab:app-ft-qwen}
\scriptsize
\setlength{\tabcolsep}{2.5pt}
\resizebox{\textwidth}{!}{%
\begin{tabular}{@{}llrrrr!{\hspace{10pt}}rrrr!{\hspace{10pt}}rrrr@{}}
\toprule
$k$ & Routers & \multicolumn{4}{c}{\textbf{\methodnaive{}}} & \multicolumn{4}{c}{\textbf{\methodmulti{}}} & \multicolumn{4}{c}{\textbf{\methodagent{}}} \\
~ & ~ & Acc & Tok & $t_\mathrm{Run}$ & $R_\mathrm{LLM}$ & Acc & Tok & $t_\mathrm{Run}$ & $R_\mathrm{LLM}$ & Acc & Tok & $t_\mathrm{Run}$ & $R_\mathrm{LLM}$ \\
\midrule
4 & 20 &\accVL{22\%}&\tokVL{23K}&\rtVL{41s}&\rndVL{1.0}&\accH{98\%}&\tokVL{29K}&\rtVL{64s}&\rndVL{3.5}&\accH{100\%}&\tokM{208K}&\rtVL{424s}&\rndM{18.8}\\
8 & 80 &\accVL{8\%}&\tokL{113K}&\rtVL{182s}&\rndVL{1.0}&\accM{91\%}&\tokVL{54K}&\rtVL{139s}&\rndVL{5.0}&\accH{98\%}&\tokM{118K}&\rtVL{341s}&\rndM{14.0}\\
12 & 180 &  &  &  &  &\accM{87\%}&\tokVL{89K}&\rtVL{192s}&\rndVL{4.3}&\accH{99\%}&\tokM{255K}&\rtL{646s}&\rndM{19.6}\\
16 & 320 &  &  &  &  &\accL{59\%}&\tokM{174K}&\rtVL{316s}&\rndVL{4.1}&\accH{99\%}&\tokM{230K}&\rtL{580s}&\rndM{18.1}\\
20 & 500 &  &  &  &  &\accVL{31\%}&\tokM{254K}&\rtL{860s}&\rndL{8.2}&\accH{99\%}&\tokM{243K}&\rtM{1422s}&\rndM{18.3}\\
30 & 1,125 &  &  &  &  &  &  &  &  &\accH{100\%}&\tokM{215K}&\rtM{1354s}&\rndM{18.1}\\
50 & 3,125 &  &  &  &  &  &  &  &  &\accH{99\%}&\tokM{287K}&\rtH{1689s}&\rndH{21.1}\\
80 & 8,000 &  &  &  &  &  &  &  &  &\accH{99\%}&\tokM{275K}&\rtM{1487s}&\rndH{20.9}\\
120 & 18,000 &  &  &  &  &  &  &  &  &\accH{99\%}&\tokM{214K}&\rtH{1722s}&\rndM{19.6}\\
160 & 32,000 &  &  &  &  &  &  &  &  & \multicolumn{4}{c}{Time Limit Exceeded} \\
200 & 50,000 &  &  &  &  &  &  &  &  &  &  &  &  \\
\bottomrule
\end{tabular}}
\end{table}

\begin{table}[H]
\centering
\caption{Three-agent comparison on intra-datacenter fat-tree with Llama-3.3-70B across all scales.}
\label{tab:app-ft-llama}
\scriptsize
\setlength{\tabcolsep}{2.5pt}
\resizebox{\textwidth}{!}{%
\begin{tabular}{@{}llrrrr!{\hspace{10pt}}rrrr!{\hspace{10pt}}rrrr@{}}
\toprule
$k$ & Routers & \multicolumn{4}{c}{\textbf{\methodnaive{}}} & \multicolumn{4}{c}{\textbf{\methodmulti{}}} & \multicolumn{4}{c}{\textbf{\methodagent{}}} \\
~ & ~ & Acc & Tok & $t_\mathrm{Run}$ & $R_\mathrm{LLM}$ & Acc & Tok & $t_\mathrm{Run}$ & $R_\mathrm{LLM}$ & Acc & Tok & $t_\mathrm{Run}$ & $R_\mathrm{LLM}$ \\
\midrule
4 & 20 &\accL{58\%}&\tokVL{20K}&\rtVL{43s}&\rndVL{1.0}&\accL{50\%}&\tokVL{19K}&\rtVL{50s}&\rndVL{2.3}&\accL{67\%}&\tokH{623K}&\rtM{1340s}&\rndH{34.8}\\
8 & 80 &\accVL{8\%}&\tokL{98K}&\rtVL{191s}&\rndVL{1.0}&\accL{67\%}&\tokVL{50K}&\rtVL{125s}&\rndVL{2.9}&\accM{92\%}&\tokH{526K}&\rtM{1273s}&\rndH{27.2}\\
12 & 180 &  &  &  &  &\accVL{17\%}&\tokVL{78K}&\rtVL{170s}&\rndVL{2.1}&\accL{58\%}&\tokM{237K}&\rtL{686s}&\rndM{18.7}\\
16 & 320 &  &  &  &  &\accVL{25\%}&\tokM{184K}&\rtVL{370s}&\rndVL{2.5}&\accLM{83\%}&\tokL{97K}&\rtVL{339s}&\rndM{12.6}\\
20 & 500 &  &  &  &  &  &  &  &  &\accLM{83\%}&\tokH{538K}&\rtH{2627s}&\rndH{28.3}\\
30 & 1,125 &  &  &  &  &  &  &  &  &\accLM{83\%}&\tokL{139K}&\rtM{1071s}&\rndM{14.4}\\
50 & 3,125 &  &  &  &  &  &  &  &  &\accM{92\%}&\tokM{373K}&\rtH{1999s}&\rndH{22.2}\\
80 & 8,000 &  &  &  &  &  &  &  &  &\accL{58\%}&\tokM{160K}&\rtM{1085s}&\rndM{15.4}\\
120 & 18,000 &  &  &  &  &  &  &  &  &\accL{67\%}&\tokM{376K}&\rtH{2320s}&\rndH{22.7}\\
160 & 32,000 &  &  &  &  &  &  &  &  &\accL{50\%}&\tokH{882K}&\rtH{5983s}&\rndH{34.1}\\
200 & 50,000 &  &  &  &  &  &  &  &  & \multicolumn{4}{c}{Time Limit Exceeded} \\
\bottomrule
\end{tabular}}
\end{table}

\subsection{Three-Agent Comparison on Inter-Datacenter and Compound-Change Analysis}
\label{app:results-complex}

We further present the complete scaling results of the three agentic methods with GPT-5.4 and GPT-4o under complex cases (inter-datacenter and compound changes) across all $k$, complementing the ``boundary'' discussion in \Cref{sec:compound}. Overall, \methodagent{} still has the smallest token-cost growth and reaches the largest scale among the three methods, while the other two again hit context-window or time limits at small scales. Under compound changes, the accuracy of \methodnaive{} and \methodmulti{} degrades more rapidly than that of \methodagent{}, indicating that reading raw configurations is even less capable of analyzing the complex case where two protocol attributes change on the same flow simultaneously---this further reveals the design advantage of \methodagent{}.

\begin{table}[H]
\centering
\caption{Three-agent comparison on inter-datacenter networks with GPT-5.4 across all scales.}
\label{tab:app-xdc-gpt-5-4}
\scriptsize
\setlength{\tabcolsep}{2.5pt}
\resizebox{\textwidth}{!}{%
\begin{tabular}{@{}llrrrr!{\hspace{10pt}}rrrr!{\hspace{10pt}}rrrr@{}}
\toprule
$k$ & Routers & \multicolumn{4}{c}{\textbf{\methodnaive{}}} & \multicolumn{4}{c}{\textbf{\methodmulti{}}} & \multicolumn{4}{c}{\textbf{\methodagent{}}} \\
~ & ~ & Acc & Tok & $t_\mathrm{Run}$ & $R_\mathrm{LLM}$ & Acc & Tok & $t_\mathrm{Run}$ & $R_\mathrm{LLM}$ & Acc & Tok & $t_\mathrm{Run}$ & $R_\mathrm{LLM}$ \\
\midrule
4 & 48 &\accH{100\%}&\tokVL{49K}&\rtVL{158s}&\rndVL{1.0}&\accM{94\%}&\tokVL{44K}&\rtVL{157s}&\rndVL{3.1}&\accHM{96\%}&\tokL{91K}&\rtVL{329s}&\rndL{7.6}\\
8 & 176 &\accHM{96\%}&\tokM{261K}&\rtL{810s}&\rndVL{1.0}&\accM{91\%}&\tokL{135K}&\rtL{451s}&\rndVL{3.1}&\accHM{96\%}&\tokL{97K}&\rtVL{394s}&\rndL{7.8}\\
12 & 384 &\accM{90\%}&\tokH{774K}&\rtH{2382s}&\rndVL{1.0}&\accM{91\%}&\tokM{333K}&\rtM{1059s}&\rndVL{2.8}&\accM{91\%}&\tokL{96K}&\rtVL{409s}&\rndL{7.8}\\
16 & 672 & \multicolumn{4}{c}{Context Window Overflow} &\accM{92\%}&\tokH{779K}&\rtH{2420s}&\rndVL{3.0}&\accM{90\%}&\tokL{100K}&\rtVL{418s}&\rndL{8.0}\\
20 & 1,040 &  &  &  &  & \multicolumn{4}{c}{Time Limit Exceeded} &\accM{88\%}&\tokL{98K}&\rtL{843s}&\rndL{7.9}\\
30 & 2,310 &  &  &  &  &  &  &  &  &\accM{86\%}&\tokL{103K}&\rtL{850s}&\rndL{7.9}\\
50 & 6,350 &  &  &  &  &  &  &  &  &\accLM{84\%}&\tokL{99K}&\rtL{921s}&\rndL{7.9}\\
80 & 16,160 &  &  &  &  &  &  &  &  &\accLM{77\%}&\tokL{98K}&\rtL{788s}&\rndL{7.9}\\
120 & 36,240 &  &  &  &  &  &  &  &  &\accLM{80\%}&\tokL{108K}&\rtM{1137s}&\rndL{8.3}\\
160 & 64,320 &  &  &  &  &  &  &  &  &\accLM{77\%}&\tokL{114K}&\rtH{1718s}&\rndL{8.7}\\
200 & 100,400 &  &  &  &  &  &  &  &  &\accLM{74\%}&\tokL{127K}&\rtH{3012s}&\rndL{9.2}\\
\bottomrule
\end{tabular}}
\end{table}

\begin{table}[H]
\centering
\caption{Three-agent comparison on inter-datacenter networks with GPT-4o across all scales.}
\label{tab:app-xdc-gpt-4o}
\scriptsize
\setlength{\tabcolsep}{2.5pt}
\resizebox{\textwidth}{!}{%
\begin{tabular}{@{}llrrrr!{\hspace{10pt}}rrrr!{\hspace{10pt}}rrrr@{}}
\toprule
$k$ & Routers & \multicolumn{4}{c}{\textbf{\methodnaive{}}} & \multicolumn{4}{c}{\textbf{\methodmulti{}}} & \multicolumn{4}{c}{\textbf{\methodagent{}}} \\
~ & ~ & Acc & Tok & $t_\mathrm{Run}$ & $R_\mathrm{LLM}$ & Acc & Tok & $t_\mathrm{Run}$ & $R_\mathrm{LLM}$ & Acc & Tok & $t_\mathrm{Run}$ & $R_\mathrm{LLM}$ \\
\midrule
4 & 48 &\accH{100\%}&\tokVL{49K}&\rtVL{57s}&\rndVL{1.0}&\accM{95\%}&\tokVL{80K}&\rtVL{118s}&\rndVL{5.2}&\accM{88\%}&\tokM{148K}&\rtVL{239s}&\rndM{12.8}\\
8 & 176 & \multicolumn{4}{c}{Context Window Overflow} &\accM{90\%}&\tokM{234K}&\rtVL{301s}&\rndVL{5.0}&\accLM{74\%}&\tokM{146K}&\rtVL{313s}&\rndM{13.1}\\
12 & 384 &  &  &  &  & \multicolumn{4}{c}{Context Window Overflow} &\accLM{72\%}&\tokM{148K}&\rtVL{348s}&\rndM{13.2}\\
16 & 672 &  &  &  &  &  &  &  &  &\accL{59\%}&\tokM{210K}&\rtVL{447s}&\rndM{16.5}\\
20 & 1,040 &  &  &  &  &  &  &  &  &\accL{52\%}&\tokM{220K}&\rtM{1381s}&\rndM{16.8}\\
30 & 2,310 &  &  &  &  &  &  &  &  &\accVL{39\%}&\tokM{304K}&\rtH{1686s}&\rndH{20.4}\\
50 & 6,350 &  &  &  &  &  &  &  &  &\accVL{43\%}&\tokH{443K}&\rtH{2424s}&\rndH{25.2}\\
80 & 16,160 &  &  &  &  &  &  &  &  &\accVL{38\%}&\tokH{463K}&\rtH{2103s}&\rndH{26.2}\\
120 & 36,240 &  &  &  &  &  &  &  &  &\accVL{37\%}&\tokH{535K}&\rtH{3373s}&\rndH{29.2}\\
160 & 64,320 &  &  &  &  &  &  &  &  & \multicolumn{4}{c}{Time Limit Exceeded} \\
200 & 100,400 &  &  &  &  &  &  &  &  &  &  &  &  \\
\bottomrule
\end{tabular}}
\end{table}

\begin{table}[H]
\centering
\caption{Three-agent comparison on compound changes with GPT-5.4 across all scales.}
\label{tab:app-cp-gpt-5-4}
\scriptsize
\setlength{\tabcolsep}{2.5pt}
\resizebox{\textwidth}{!}{%
\begin{tabular}{@{}llrrrr!{\hspace{10pt}}rrrr!{\hspace{10pt}}rrrr@{}}
\toprule
$k$ & Routers & \multicolumn{4}{c}{\textbf{\methodnaive{}}} & \multicolumn{4}{c}{\textbf{\methodmulti{}}} & \multicolumn{4}{c}{\textbf{\methodagent{}}} \\
~ & ~ & Acc & Tok & $t_\mathrm{Run}$ & $R_\mathrm{LLM}$ & Acc & Tok & $t_\mathrm{Run}$ & $R_\mathrm{LLM}$ & Acc & Tok & $t_\mathrm{Run}$ & $R_\mathrm{LLM}$ \\
\midrule
4 & 20 &\accH{100\%}&\tokVL{20K}&\rtVL{66s}&\rndVL{1.0}&\accH{100\%}&\tokVL{30K}&\rtVL{107s}&\rndVL{3.2}&\accH{100\%}&\tokVL{48K}&\rtVL{178s}&\rndVL{6.4}\\
8 & 80 &\accLM{80\%}&\tokL{100K}&\rtVL{316s}&\rndVL{1.0}&\accH{97\%}&\tokVL{57K}&\rtVL{201s}&\rndVL{3.1}&\accH{99\%}&\tokVL{55K}&\rtVL{227s}&\rndVL{6.6}\\
12 & 180 &\accLM{70\%}&\tokM{293K}&\rtL{904s}&\rndVL{1.0}&\accM{87\%}&\tokL{123K}&\rtVL{410s}&\rndVL{3.0}&\accH{97\%}&\tokVL{55K}&\rtVL{240s}&\rndVL{6.3}\\
16 & 320 &\accL{65\%}&\tokH{645K}&\rtH{1978s}&\rndVL{1.0}&\accM{88\%}&\tokM{249K}&\rtL{792s}&\rndVL{3.0}&\accHM{96\%}&\tokVL{58K}&\rtVL{245s}&\rndVL{6.4}\\
20 & 500 & \multicolumn{4}{c}{Context Window Overflow} &\accL{68\%}&\tokH{440K}&\rtH{1510s}&\rndVL{2.9}&\accM{95\%}&\tokVL{57K}&\rtL{532s}&\rndVL{6.3}\\
30 & 1,125 &  &  &  &  & \multicolumn{4}{c}{Time Limit Exceeded} &\accLM{82\%}&\tokVL{57K}&\rtL{469s}&\rndVL{6.3}\\
50 & 3,125 &  &  &  &  &  &  &  &  &\accM{80\%}&\tokVL{61K}&\rtL{571s}&\rndVL{6.5}\\
80 & 8,000 &  &  &  &  &  &  &  &  &\accLM{78\%}&\tokVL{57K}&\rtL{513s}&\rndVL{6.3}\\
120 & 18,000 &  &  &  &  &  &  &  &  &\accLM{76\%}&\tokVL{57K}&\rtL{750s}&\rndVL{6.3}\\
160 & 32,000 &  &  &  &  &  &  &  &  &\accLM{78\%}&\tokVL{58K}&\rtM{1179s}&\rndVL{6.3}\\
200 & 50,000 &  &  &  &  &  &  &  &  &\accLM{79\%}&\tokVL{58K}&\rtH{2100s}&\rndVL{6.3}\\
\bottomrule
\end{tabular}}
\end{table}

\begin{table}[H]
\centering
\caption{Three-agent comparison on compound changes with GPT-4o across all scales.}
\label{tab:app-cp-gpt-4o}
\scriptsize
\setlength{\tabcolsep}{2.5pt}
\resizebox{\textwidth}{!}{%
\begin{tabular}{@{}llrrrr!{\hspace{10pt}}rrrr!{\hspace{10pt}}rrrr@{}}
\toprule
$k$ & Routers & \multicolumn{4}{c}{\textbf{\methodnaive{}}} & \multicolumn{4}{c}{\textbf{\methodmulti{}}} & \multicolumn{4}{c}{\textbf{\methodagent{}}} \\
~ & ~ & Acc & Tok & $t_\mathrm{Run}$ & $R_\mathrm{LLM}$ & Acc & Tok & $t_\mathrm{Run}$ & $R_\mathrm{LLM}$ & Acc & Tok & $t_\mathrm{Run}$ & $R_\mathrm{LLM}$ \\
\midrule
4 & 20 &\accLM{78\%}&\tokVL{20K}&\rtVL{25s}&\rndVL{1.0}&\accLM{84\%}&\tokVL{29K}&\rtVL{45s}&\rndVL{3.3}&\accLM{81\%}&\tokL{98K}&\rtVL{160s}&\rndM{12.5}\\
8 & 80 &\accVL{33\%}&\tokL{100K}&\rtVL{111s}&\rndVL{1.0}&\accVL{30\%}&\tokVL{59K}&\rtVL{90s}&\rndVL{3.3}&\accLM{71\%}&\tokL{108K}&\rtVL{229s}&\rndM{13.2}\\
12 & 180 & \multicolumn{4}{c}{Context Window Overflow} &\accVL{19\%}&\tokL{131K}&\rtVL{171s}&\rndVL{3.3}&\accVL{42\%}&\tokL{112K}&\rtVL{268s}&\rndM{13.3}\\
16 & 320 &  &  &  &  &\accVL{17\%}&\tokM{257K}&\rtVL{295s}&\rndVL{3.2}&\accVL{44\%}&\tokL{114K}&\rtVL{258s}&\rndM{13.3}\\
20 & 500 &  &  &  &  & \multicolumn{4}{c}{Context Window Overflow} &\accL{52\%}&\tokL{119K}&\rtL{887s}&\rndM{13.5}\\
30 & 1,125 &  &  &  &  &  &  &  &  &\accVL{38\%}&\tokM{173K}&\rtL{899s}&\rndM{15.4}\\
50 & 3,125 &  &  &  &  &  &  &  &  &\accVL{38\%}&\tokM{206K}&\rtM{1225s}&\rndM{17.2}\\
80 & 8,000 &  &  &  &  &  &  &  &  &\accVL{40\%}&\tokM{220K}&\rtM{1169s}&\rndM{17.5}\\
120 & 18,000 &  &  &  &  &  &  &  &  &\accVL{36\%}&\tokM{292K}&\rtH{2054s}&\rndM{19.1}\\
160 & 32,000 &  &  &  &  &  &  &  &  &\accVL{41\%}&\tokM{276K}&\rtH{3272s}&\rndM{18.9}\\
200 & 50,000 &  &  &  &  &  &  &  &  & \multicolumn{4}{c}{Time Limit Exceeded} \\
\bottomrule
\end{tabular}}
\end{table}

\section{Discussion}
\label{app:discussion}

\paragraph{Topology coverage.}
Production datacenter networks vary in subtle layer counts and structural details across vendors---Azure~\cite{azure-dc}, Google's Jupiter~\cite{jupiter}, and Meta's fabric~\cite{fb-fabric} all differ in how many tiers they expose and how pods are aggregated (e.g., Meta's fabric has three tiers and within pods 48 racks are aggregated by four fabric switches~\cite{fb-fabric}, while Azure deploys three tiers (T0/T1/T2) with T0 aggregated under T1 to form pods and pods interconnected by T2 spines~\cite{azure-dc}), but they are all Clos networks at the core. 
To reflect a generic production datacenter, we adopt fat-trees, which are widely used as the reference Clos topology in academia~\cite{fattree1,fattree2} and as the building block of production datacenter fabrics~\cite{azure-dc,jupiter,fb-fabric}. 
Supporting other Clos variants (leaf-spine~\cite{azure-dc}, dragonfly~\cite{dragonfly}) requires only new topology templates in our generation pipeline.

\paragraph{Configuration coverage.}
\benchmarkname{} models routing-related configurations only (\Cref{sec:benchmark}) to construct a pure benchmark for routing-path analysis.
Real networks also contain non-routing configurations, e.g., security policies, traffic bandwidth-allocation rules, device login-authentication settings, and device/traffic status-monitoring configurations~\cite{acl-cisco}. 
\Cref{app:results-realnet} shows that this heterogeneity may inflate token cost on the production network. 
We hope future agentic-NetOps benchmarks will incorporate and analyze these protocols beyond path analysis.

\paragraph{Cost measurement under remote LLM APIs.}
LLM API runtimes can be perturbed by remote-service load and network conditions, making runtime a noisy cost signal.
We therefore use \emph{token consumption} as the primary cost metric for each agent, which remains stable regardless of network conditions and workloads. 
Additionally, we pin every LLM to a specific OpenAI API version or open-source model release (\Cref{sec:results}) to minimize result drift across versions of the same model.

\end{CJK*}

\end{document}